\documentclass{article}

\pdfoutput=1
\usepackage{jheppub}
\usepackage{hyperref}
\usepackage{datetime}
\usepackage{float}

\usepackage{amsmath}
\usepackage{amssymb}
\usepackage{amsthm}
\usepackage{mathrsfs}
\usepackage{slashed}
\usepackage{tabularx}

\def\ttt{\texttt}

\def\q{\,q}

\def\C{\mathbb{C}}
\def\Q{\mathbb{Q}}
\def\R{\mathbb{R}}
\def\Z{\mathbb{Z}}
\def\bA{\mathbb{A}}

\def\bB{\mathbb{B}}
\def\bT{\mathbb{T}}

\def\cA{\mathcal{A}}
\def\cB{\mathcal{B}}
\def\cC{\mathcal{C}}
\def\cD{\mathcal{D}}
\def\cE{\mathcal{E}}
\def\cF{\mathcal{F}}
\def\cG{\mathcal{G}}
\def\cH{\mathcal{H}}
\def\cI{\mathcal{I}}
\def\cJ{\mathcal{J}}
\def\cK{\mathcal{K}}
\def\cL{\mathcal{L}}
\def\cM {\mathcal{M}}
\def\cN{\mathcal{N}}
\def\cO{\mathcal{O}}
\def\cP {\mathcal{P}}
\def\cQ {\mathcal{Q}}
\def\cR{\mathcal{R}}
\def\cS{\mathcal{S}}
\def\cT{\mathcal{T}}
\def\cU{\mathcal{U}}
\def\cV{\mathcal{V}}
\def\cW{\mathcal{W}}
\def\cX{\mathcal{X}}
\def\cY {\mathcal{Y}}
\def\cZ{\mathcal{Z}}

\def\dg {\dagger}
\def\p{\partial}
\def\sp {\slashed{\p}}
\def\sD{\slashed{D}}

\def\/{\over}

\def\rn{\rangle}
\def\ln{\langle}

\def\t{\theta}
\def\s{\sigma}
\def\e{\epsilon}
\def\ve{\varepsilon}
\def\vphi{\varphi}
\def\a{\alpha}
\def\b{\beta}
\def\d{\delta}
\def\k{\kappa}
\def\g {\gamma}
\def\la {\lambda}
\def\w {\omega}
\def\z{\zeta}
\def\u {\mu}
\def\l{\ell}
\def\mn{{\mu\nu}}
\def\rs{{\rho\sigma}}
\def\ab{{\a\b}}

\def\n {\nabla}
\def\L{\Lambda}
\def\D{\Delta}
\def\G {\Gamma}
\def\Om {\Omega}

\def\ra{\rightarrow}
\def\lra{\longrightarrow}

\def\Tr{\mathrm{Tr}}
\def\Diag{\mathrm{Diag}}
\def\Str{\mathrm{Str}}
\def\sdim{\mathrm{sdim}}
\def\Ker{\mathrm{Ker}}
\def\Vol{\mathrm{Vol}}
\def\SUSY{\text{SUSY}}
\def\BH{\text{BH}}
\def\vac{\text{vac}}
\def\Im{\mathrm{Im}\,}
\def\Re	{\mathrm{Re}\,}
\def\r{\mathrm}
\def\hc{\text{h.c.}}

\def\_{\hspace{2cm}}
\def\'{\:\:}

\newcommand{\bea}{\begin{eqnarray}}
	\newcommand{\eea}{\end{eqnarray}}
\def\({\left(}
\def\){\right)}

\def\tF{\tilde{F}}
\def\bpsi{\bar{\psi}}
\def\bchi{\bar{\chi}}

\newcommand{\be}{\begin{equation}}
\newcommand{\ee}{\end{equation}}
\def\le{\left}
\def\ri{\right}

    \author[a]{Alejandra Castro,} \author[a]{Victor Godet,} \author[b]{Finn Larsen} \author[b]{and Yangwenxiao Zeng}
    \title{Logarithmic Corrections to Black Hole Entropy: \\the Non-BPS Branch}
    \affiliation[a]{Institute for Theoretical Physics Amsterdam and Delta Institute for Theoretical Physics, University of Amsterdam, Science Park 904, 1098 XH Amsterdam, the Netherlands.}
    \affiliation[b]{Department of Physics and Leinweber Center for Theoretical Physics, University of Michigan, 450 Church Street, Ann Arbor, MI 48109-1120, USA.}

 \abstract{We compute the leading logarithmic correction to black hole entropy on the non-BPS branch of 4D ${\cal N}\geq 2$ supergravity theories.
This branch corresponds to finite temperature black holes whose extremal limit does not preserve supersymmetry, such as the $D0-D6$ system in string theory.
Starting from a black hole in minimal Kaluza-Klein theory, we discuss in detail its embedding into
${\cal N}=8, 6, 4, 2$ supergravity, its spectrum of quadratic fluctuations in all these environments, and the resulting quantum corrections.
We find that the $c$-anomaly vanishes only when ${\cal N}\geq 6$, in contrast to the BPS branch where $c$ vanishes for all ${\cal N}\geq 2$.
We briefly discuss potential repercussions  this feature could have in a microscopic description of these black holes.}

\begin{document}
    \maketitle

    \section{Introduction}

A remarkable feature of the Bekenstein-Hawking entropy formula is its universality: the leading contribution to the black hole entropy is controlled by the area of the event horizon, regardless of the details of the solutions or the matter content of the theory. It is therefore interesting to investigate if there is any notion of universality and/or robustness in the quantum corrections to the entropy of a black hole.

Generically there is no expectation that the quantum corrections to the Bekenstein-Hawking area law are universal: according to effective quantum field theory they are sensitive to the details of the UV completion of the low energy theory in consideration. However, there is a special class of quantum corrections that are entirely determined by the low energy theory \cite{BanerjeeGuptaMandalEtAl2011,Banerjee2011,Sen2012b,Sen:2012cj,Chowdhury:2014lza,Gupta:2014hxa,Keeler:2014bra,Keeler2014,LarsenLisbao2015}: the leading {\it logarithmic} correction is governed by the one-loop effective action of the low energy modes in the gravitational theory. These corrections, therefore, provide a powerful \textit{infrared window into the microstates}.

The claim that logarithmic corrections computed from the IR theory agree with results for the UV completion has been successfully tested in many cases where string theory provides a microscopic counting formula for black hole microstates. We refer to \cite{Mandal:2010cj,Sen2014} for a broad overview and \cite{BhattacharyyaGrassiMarinoEtAl2014,JeonLal2017,LiuPandoZayasRatheeEtAl2017,LiuPandoZayasRatheeEtAl2017a} for more recent developments
in AdS$_4$/CFT$_3$.  Logarithmic corrections have also been evaluated for a plethora of other black holes \cite{Sen:2012dw,Charles:2015eha} where
a microscopic account still awaits.\footnote{In certain cases the logarithm can be accounted for very simply by using thermodynamics \cite{Sen:2012dw,Pathak2017}: the measure that controls the change from, for example, the microcanonical to the canonical ensemble correctly reproduces the gravitational result without leading to new insight in the microscopic theory.}

The coefficients multiplying these logarithms follow some interesting patterns.
The black hole entropy has the schematic structure
\be\label{eq:sbhlog(intro)}
S_{\rm BH}={A_{H}\over 4G} + {1\over 2}(C_{\rm local} +C_{\rm zm}) \log {A_H\over G} + \cdots~,
\ee
where we highlight the two terms (area law+logarithm) controlled by low energy gravity and use dots to denote subleading corrections that generally
depend on the UV completion.
$C_{\rm zm}$ is an integer that accounts for zero modes in the path integral. $C_{\rm local}$ refers to the constant term in the heat kernel that
captures the non-zero eigenvalues of the one-loop determinant \cite{Hawking1977}. It is expressed as a density \cite{Birrell:1982ix,Vassilevich2003}
\be\label{eq:clocal}
C_{\rm local}= \int d^4x \sqrt{g}\, a_4(x)~,
\ee
where the integrand takes the form
\be
a_4(x)= {c\over 16\pi^2} W_{\mu\nu\rho\sigma}W^{\mu\nu\rho\sigma}-{a\over 16\pi^2} E_4~,
\ee
for the backgrounds we will consider.
In this expression, $E_4$ is the Gauss-Bonnet term and $W_{\mu\nu\rho\sigma}$ is the Weyl tensor. The constants $c$ and $a$ are familiar from related computations of
the trace anomaly of the stress tensor. Their values depend on the content of matter fields and their couplings to the background black hole solution.\footnote{It is important to note that the couplings are not necessarily minimal, so the values of $c$ and $a$ may be nonstandard functions of the matter content.}

$C_{\rm local}$ contains non-trivial information about the background so this function generally depends greatly on the matter content of the theory and the parameters of the black hole \cite{Sen:2012dw}. However, under certain conditions $C_{\rm local}$ has a universal structure \cite{Charles:2015eha,Charles2017}: for Kerr-Newman black holes embedded in ${\cal N}\geq2$ supergravity, the $c$-anomaly vanishes. This leads to a remarkable simplification since then the integral in \eqref{eq:clocal} is just a topological
invariant. The logarithmic correction is therefore universal in the sense that its does not depend on details of the black hole background; it is determined entirely
by the content of massless fields.

The class of backgrounds considered in \cite{Charles:2015eha} was constructed such that, in the extremal limit, they continuously connect to BPS solutions.
For this reason we denote this class as the \textit{BPS branch}. The black holes on the BPS branch are not generally supersymmetric, but their couplings to matter
are arranged such that supersymmetry is attained in the limit. One of the motivations for the present article is to study universality of logarithmic corrections outside of the BPS branch in $D=4$ supergravity.

Supergravity (with ${\cal N}\geq2$) also allows for black holes that do not approach BPS solutions in the extremal limit. We refer to such solutions as the \textit{non-BPS branch}. In their minimal incarnation, they correspond to solutions of the $D=4$ theory obtained by a Kaluza-Klein reduction of five dimensional
Einstein gravity \cite{Larsen:1999pp}. In a string theory setup it is natural to identify the compact Kaluza-Klein dimension with the M-theory circle, and
then these solutions are charged with respect to electric $D0$-brane charge and magnetic $D6$-brane charge. Such configurations break supersymmetry even in the extremal limit. Therefore, they offer an interesting arena for studying logarithmic corrections and their possible universality.

The minimal Kaluza-Klein theory needed to describe the non-BPS branch is a four dimensional Einstein-Maxwell-dilaton theory where the couplings are dictated by the reduction from five dimensions. We will refer to the black hole solutions of this theory as ``Kaluza-Klein black holes.'' These solutions can be embedded in supergravity, as we will discuss in detail. In particular, we will consider the embedding of the Kaluza-Klein theory in  ${\cal N}=4,6,8$ supergravity and for ${\cal N}=2$ we
consider $ST(n)$ models \footnote{We work out the bosonic fluctuations for ${\cal N}=2$ with any prepotential. It is only for fermionic fluctuations that we restrict our attention to the $ST(n)$ models.}, which include the well-known $STU$-model as a special case.

Our technical goal is to evaluate the Seeley-DeWitt coefficient $a_4(x)$ for the Kaluza-Klein black hole when it is embedded in one of the supergravities. This involves the study of quadratic fluctuations around the background, potentially a formidable task since there are many fields and generally they have non-minimal couplings to the
background and to each other. Fortunately we find that, in the cases we consider, global symmetries of supergravity organize the quadratic fluctuations
into manageable groups of fields that are decoupled from one another.
We refer to such groups of fields as ``blocks". There are only five distinct types of blocks, summarized in Table \ref{IntroTabel}.
 \begin{table}[H]
	\centering\renewcommand{\arraystretch}{1.3}
\begin{tabular}{|c|c|c|}
	\hline
	Multiplet 	& Block content  \\\hline\hline
 KK block & 1 graviton, 1 vector, 1 scalar \\\hline
	Vector block  & 1 vector and 1 (pseudo)scalar  \\\hline
 Scalar block & 1 real scalar \\\hline
	Gravitino block &2 gravitini and 2 gaugini   \\\hline
	Gaugino block &2 gaugini \\
	\hline
\end{tabular}
	
\caption{Decomposition of quadratic fluctuations.}\label{IntroTabel}
\end{table}
\noindent
The KK block comprises the quadratic fluctuations in the seed theory, {\it i.e.} the Kaluza Klein theory with no additional matter fields. The scalar block is a single minimally coupled spectator scalar field. The remaining matter blocks have unfamiliar field content and their couplings to the background are non-standard.
The great simplification is that the spectrum of quadratic fluctuations of each supergravity theory we consider can be characterized by the number of times
each type of block appears. We record those degeneracies in Tables \ref{tab:degeneracy} and \ref{tab:degeneracy1}.

Once the relevant quadratic fluctuations  are identified it is a straightforward (albeit cumbersome) task to evaluate the Seeley-DeWitt coefficient $a_4(x)$. We do this for every block listed above
and so determine their contribution to $C_{\rm local}$ in \eqref{eq:sbhlog(intro)}. Having already computed the degeneracies of the blocks, it is elementary algebra to find the values of $c$ and $a$ for each supergravity theory. Our results for individual blocks are given in Table \ref{KKres} and those for theories are given in
Table \ref{tab:degeneracy1}.

One of our main motivation is to identify theories where $c=0$ since for those the coefficient of the logarithm is universal. We find that the non-trivial cancellations on the BPS branch reported in \cite{Charles:2015eha} are much rarer on the non-BPS branch. For example, on the non-BPS branch the $c$ coefficient does not vanish for
any ${\cal N}=2,4$ supergravity we consider, whatever their matter content. Therefore, as we discuss in section \ref{sec:lognonbps}, this implies that the logarithmic correction to the entropy depends on black hole parameters in a combination different from the horizon area.

In contrast, for ${\cal N}=6,8$ we find that $c=0$. The vanishing of $c$ on the non-BPS branch is rather surprising, since it is apparently due to a different balance among the field content and couplings than the analogous cancellation on the BPS-branch. It would be very interesting to understand the origin of this cancellation from a more fundamental principle. In our closing remarks we discuss some directions to pursue.

The outline of this paper is as follows. In section \ref{sec:KK theory}, we discuss Kaluza-Klein theory and its Kaluza-Klein black hole solution. This gives the ``seed solution", the minimal incarnation of the non-BPS branch. In section \ref{sec:kkn8}, we embed this theory into ${\cal N}=8$ supergravity, and in section \ref{sec:fluctuations}, we derive the quadratic fluctuations around the black hole in the ${\cal N}=8$ environment. In section \ref{sec:n8}, we discuss the embedding of the Kaluza-Klein
black hole into theories with less supersymmetry by truncating our previous results for ${\cal N}=8$ and then exploiting global symmetries of supergravity.
In section \ref{sec:N=2sugra}, we discuss the embedding of the non-BPS branch directly into ${\cal N}=2$ supergravity, without making reference to ${\cal N}=8$. This generalizes some ${\cal N}=2$ results to a general prepotential. In section \ref{sec:lognonbps}, we evaluate the $c$ and $a$ coefficients for the Kaluza-Klein black hole in its various embeddings and discuss the resulting quantum corrections to the black hole entropy. Finally, section \ref{sec:discussion} summarizes our results and discusses future directions.
Appendix \ref{app:a4} contains the technical details behind the Seeley-DeWitt coefficients presented in section \ref{sec:lognonbps}.

 \section{The Kaluza-Klein Black Hole}
 \label{sec:KK theory}

Our starting point is a black hole solution to Kaluza-Klein theory. It is sufficient for our purposes to consider the original version of Kaluza-Klein theory: the compactification to four spacetime dimensions of Einstein gravity in five dimensions. In this section, we briefly present the theory and its
black hole solutions. In the following sections we embed the theory and its solutions into supergravity and study perturbations around the
Kaluza-Klein black holes in the framework
of supergravity.

\medskip

The Lagrangian of Kaluza-Klein theory is given by\footnote{We use $e$ and $\sqrt{-g}$ interchangeably, to denote the square root of the determinant of the metric.}
\begin{align}
e^{-1}\mathcal{L}_{\text{KK}}=\frac{1}{16\pi G}\left(R-2 D_{\mu}\Phi D^{\mu}\Phi-\frac{1}{4}e^{-2\sqrt{3}\Phi}F_{\mu\nu}F^{\mu\nu}\right)~.
\label{eqn:KKtheory}
\end{align}
The scalar field $\Phi$ parametrizes the size of the compact fifth dimension and the field strength $F_{\mu\nu}$ is the 4D
remnant of the metric with one index along the fifth dimension. The Lagrangian (\ref{eqn:KKtheory}) gives the equations of motion
\begin{align}\label{EoM:PhiKK}
& D^{2}\Phi+\frac{\sqrt{3}}{8}e^{-2\sqrt{3}\Phi}F_{\mu\nu}F^{\mu\nu}=0~,\\
\label{EoM:AKK}
& D_{\mu}\left(e^{-2\sqrt{3}\Phi}F^{\mu\nu}\right)=0~,\\
\label{EoM:gKK}
&R_{\mu\nu}-\frac{1}{2}g_{\mu\nu}R=\left(2 D_{\mu}\Phi D_{\nu}\Phi-g_{\mu\nu} D^{\rho}\Phi D_{\rho}\Phi\right)
+\frac{1}{2}e^{-2\sqrt{3}\Phi}\left(F_{\mu\rho}F_{ \nu}^{\ \rho}-\frac{1}{4}g_{\mu\nu}F_{\rho\sigma}F^{\rho\sigma}\right)~.
\end{align}

Some of our considerations will apply to any solution of the Kaluza-Klein theory (\ref{eqn:KKtheory}) but our primary interest is in
asymptotically flat black holes. We therefore focus on the general Kaluza-Klein black
hole \cite{Rasheed:1995zv,Larsen:1999pp,Horowitz:2011cq}.
It is characterized by the black hole mass $M$ and angular momentum $J$, along with the electric/magnetic charges $(Q,P)$ of the
Maxwell field. Its 4D metric is given by
\begin{align}
ds^{2}_{4}= g^{(\text{KK})}_{\mu\nu} dx^\mu dx^\nu =
-\frac{H_3}{\sqrt{H_{1}H_{2}}}\left(dt-B\right)^{2}+\sqrt{H_{1}H_{2}}\left(\frac{dr^{2}}{\Delta}+d\theta^{2}+\frac{\Delta}{H_3}\r{sin}^{2}\theta \,d\phi^{2}\right)~,
\label{eqn:metric}
\end{align}
where
\begin{eqnarray}
H_{1}&=&r^{2}+a^{2}\r{cos}^{2}\theta+r(p-2m)+\frac{p}{p+q}\frac{(p-2m)(q-2m)}{2}\nonumber\\
&&-\frac{p}{2m(p+q)}\sqrt{(q^{2}-4m^{2})(p^{2}-4m^{2})}a\cos\theta~,\\
H_{2} &=&r^{2}+a^{2}\r{cos}^{2}\theta+r(q-2m)+\frac{q}{p+q}\frac{(p-2m)(q-2m)}{2}\nonumber\\
&&+\frac{q}{2m(p+q)}\sqrt{(q^{2}-4m^{2})(p^{2}-4m^{2})}a\cos\theta~,\\
H_3 &=&r^{2}-2mr+a^{2}\r{cos}^{2}\theta~,\\
\Delta &=&r^{2}-2mr+a^{2}~,
\end{eqnarray}
and the 1-form $B$ is given by
\begin{align}
B &=\sqrt{pq}\frac{(pq+4m^{2})r-m(p-2m)(q-2m)}{2m(p+q)H_3}a\,\r{sin}^{2}\theta\, d\phi~.
\end{align}
The matter fields are the gauge field
\begin{eqnarray}
A^{(\text{KK})} &=&-\left[2Q\left(r+\frac{p-2m}{2}\right)+\sqrt{\frac{q^{3}(p^{2}-4m^{2})}{4m^{2}(p+q)}}a\cos\theta\right]H_{2}^{-1}dt\nonumber\\
&&-\left[2P\left(H_{2}+a^{2}\sin^{2}\theta\right)\cos\theta+\sqrt{\frac{p(q^{2}-4m^{2})}{4m^{2}(p+q)^{3}}}\right.\nonumber\\
&& \left. \times \vphantom{\sqrt{\frac{p(q^{2})}{4m^{2}(p+q)^{3}}}} \left[(p+q)(pr-m(p-2m))+q(p^{2}-4m^{2})\right]a\,\r{sin}^{2}\theta\right]H_{2}^{-1}d\phi~,
\label{eqn:KKgauge}
\end{eqnarray}
and the dilaton
\begin{align}
e^{-4\Phi^{(\text{KK})}/ \sqrt{3}}=\sqrt{\frac{H_{2}}{H_{1}}}~.
\end{align}
The superscript ``KK" on $g^{(\text{KK})}_{\mu\nu}$, $A^{(\text{KK})}$ , and $\Phi^{(\text{KK})}$ refers to the Kaluza-Klein black hole.
These background fields should be distinguished from the exact fields in (\ref{eqn:KKtheory}-\ref{EoM:gKK}) which generally include fluctuations
around the background.

The four parameters $m,a,p,q$ appearing in the solution determine the four physical parameters $M,J,Q,P$ as
\begin{align}
2GM & =\frac{p+q}{2}~,\\
GJ &=\frac{\sqrt{pq}(pq+4m^{2})}{4(p+q)} {a\over m}~,\\
Q^{2} &=\frac{q(q^{2}-4m^{2})}{4(p+q)}~,\\
P^{2} &=\frac{p(p^{2}-4m^{2})}{4(p+q)}~.
\end{align}
Note that $q,p\geq2m$, with equality corresponding to the absence of electric or magnetic charge, respectively.

The spectrum of quadratic fluctuations around the general black hole solution to Kaluza-Klein theory is
complicated. In section \ref{sec:N=2sugra} we start with a
general solution to the equations of motion (\ref{EoM:PhiKK}-\ref{EoM:gKK})
such as the Kaluza-Klein black hole $g^{(\text{KK})}_{\mu\nu}$, $A^{(\text{KK})}_{\mu}$, and $\Phi^{(\text{KK})}$ presented above.
We construct an embedding into $\mathcal{N}=2$ SUGRA with arbitrary cubic prepotential and study fluctuations around the background.
Although we make some progress in this general setting it proves notable that the analysis simplifies greatly when the
background dilaton is constant $\Phi^{(\text{KK})}=0$.

In the predominant part of the paper we therefore focus on the simpler case from the outset and assume $\Phi^{(\text{KK})}=0$.
We arrange this by considering the non-rotating black hole $J=0$ with $P^2=Q^2$. In this special case the
metric $g^{(\text{KK})}_{\mu\nu}$ is (\ref{eqn:metric}) with
\begin{align}
H_1 & = H_2  = \left( r+\frac{q-2m}{2} \right)^2 ~,\cr
H_3 & = \Delta = r^{2}-2mr~,
\label{eqn:KKh123}
\end{align}
and the gauge field (\ref{eqn:KKgauge}) becomes
\begin{align}
A^{(\text{KK})}  = - 2Q\left(r+\frac{q-2m}{2}\right)^{-1}dt  -2 P \cos\theta d\phi~.
\label{eqn:KKAKK}
\end{align}
In the simplified setting it is easy to eliminate the parameters $m,q$ in favor of the physical mass $2GM=q$ and charges
$P^2 = Q^2 = {1\over 8} ( q^2 - 4m^2)$ but we do not need to do so.

When $\Phi^{(\text{KK})}=0$ the geometry of the Kaluza-Klein black hole is in fact the same as the Reissner-Nordstr\"{o}m
black hole. Indeed, they both satisfy the standard Einstein-Maxwell equations
\begin{align}
\label{EoM:gKK2}
&R^{(\text{KK})}_{\mu\nu}=\frac{1}{2} \left( F^{(\text{KK})}_{\mu\rho}F_{ \nu}^{(\text{KK}) \rho} - \frac{1}{4} g_{\mu\nu}
F^{(\text{KK})}_{\rho\sigma} F^{(\text{KK})\rho\sigma} \right)~,\\
\label{EoM:AKK2}
& D_{\mu}F^{(\text{KK})\mu\nu}=0~.
\end{align}
However, whereas the Reissner-Nordstr\"{o}m solution can be supported by any combination of
electric and magnetic charges $(Q,P)$ with the appropriate value of $Q_{\rm eff} = \sqrt{P^2 + Q^2}$, for the Kaluza-Klein black hole
we must set $P^2=Q^2$ so
\begin{align}
\label{EoM:PhiKK2}
F^{(\text{KK})}_{\mu\nu}F^{(\text{KK})\mu\nu}=0~,
\end{align}
or else the dilaton equation of motion (\ref{EoM:PhiKK}) is inconsistent with a constant dilaton $\Phi^{(\text{KK})}$.
This difference between the two cases is closely related to the fact that, after embedding in supergravity, the Kaluza-Klein black hole does not preserve supersymmetry in the extremal limit.

\section{The KK Black Hole in $\mathcal{N}=8$ SUGRA}
\label{sec:kkn8}
\label{sec:N=8SUGRA}
In this section, we review $\mathcal{N}=8$ SUGRA and show how to embed a solution of $D=4$ Kaluza-Klein theory with
constant dilaton into $\mathcal{N}=8$ SUGRA.

\subsection{$\mathcal{N}=8$ Supergravity in Four Dimensions}
The matter content of $\mathcal{N}=8$ SUGRA is a spin-2 graviton $g_{\mu\nu}$, $8$ spin-3/2 gravitini $\psi_{A\mu}$ (with $A=1,...,8$),
$28$ spin-1 vectors $B_{\mu}^{MN}$ (antisymmetric in $M,N=1,...,8$), $56$ spin-1/2 gaugini
$\lambda_{ABC}$ (antisymmetric in $A,B,C=1,...,8$), and $70$ spin-0 scalars. The Lagrangian can be
presented as \cite{Cremmer:1979up}\footnote{To match with the conventions of many authors, when discussing ${\cal N}=8$ supergravity, we set Newton constant to $\kappa^2 = 8\pi G = 2$. In section \ref{sec:N=2sugra}, we will restore the explicit  $\kappa$ dependence.}
\begin{eqnarray}\label{N=8:Lagrangian}
e^{-1}\mathcal{L}^{(\mathcal{N}=8)}&=&\frac{1}{4}R-\frac{1}{2}\bar{\psi}_{A\mu}\gamma^{\mu\nu\rho}D_{\nu}\psi_{A\rho}-\frac{i}{8}G_{\mu\nu}^{MN}\widetilde{H}_{MN}^{\text{(F)}\mu\nu}-\frac{1}{12}\bar{\lambda}_{ABC}\gamma^{\mu}D_{\mu}\lambda_{ABC}\nonumber\\
&&-\frac{1}{24}P_{\mu ABCD}\bar{P}^{\mu ABCD}-\frac{1}{6\sqrt{2}}\bar{\psi}_{A\mu}\gamma^{\nu}\gamma^{\mu}\left(\bar{P}_{\nu}^{ABCD}+\hat{\bar{P}}_{\nu}^{ABCD}\right)\lambda_{BCD}\\
&&+\frac{1}{8\sqrt{2}}\left(\bar{\psi}_{A\mu}\gamma^{\nu}\hat{\mathcal{F}}_{AB}\gamma^{\mu}\psi_{\nu B}-\frac{1}{\sqrt{2}}\bar{\psi}_{C\mu}\hat{\mathcal{F}}_{AB}\gamma^{\mu}\lambda_{ABC}+\frac{1}{72}\epsilon^{ABCDEFGH}\bar{\lambda}_{ABC}\hat{\mathcal{F}}_{DE}\lambda_{FGH}\right)~,\nonumber
\end{eqnarray}
in conventions where all fermions are in Majorana form, the metric is ``mostly plus", and Hodge duality is defined by
\begin{align}
&\widetilde{H}_{MN}^{\text{(F)}\mu\nu}=-\frac{i}{2}\epsilon^{\mu\nu\rho\sigma}H^{\text{(F)}}_{MN\rho\sigma}~,\ \epsilon_{0123}=e~.
\end{align}
Below we also use $(R/L)$ superscripts on fermions, to denote their right- and left-handed components.

We include all the glorious details of $\mathcal{N}=8$ SUGRA to facilitate comparison with other references. The symmetry structure is the most important aspect for our applications so we focus on that in the following. The starting point is the $56$-bein
\begin{align}
\mathcal{V}=\begin{pmatrix}U_{AB}^{\quad MN} & V_{ABMN} \\ \bar{V}^{ABMN} & \bar{U}^{AB}_{\quad MN}\end{pmatrix}~,
\label{eqn:56bein}
\end{align}
that is acted on from the left by a local $SU(8)$ symmetry (with indices $A,B,\ldots$) and from the right by a global $E_{7(7)}$ duality
symmetry (with indices $M,N$). The connection
\begin{align}
&\partial_{\mu}\mathcal{V}\mathcal{V}^{-1}=\begin{pmatrix}2Q_{\mu[A}^{\quad[C}\delta_{B]}^{\ D]} & P_{\mu ABCD} \\ \bar{P}_{\mu}^{ABCD} & 2\bar{Q}_{\mu\ [C}^{\ [A}\delta^{B]}_{\ D]}\end{pmatrix}~,
\end{align}
defines an $SU(8)$ gauge field $Q_{\mu A}^{\quad B}$ that renders the $SU(8)$ redundant. We therefore interpret $P_{\mu ABCD}$
as covariant derivatives of scalar fields that belong to the coset $E_{7(7)}/SU(8)$ with dimension $133-63=70$. The term
in (\ref{N=8:Lagrangian}) that is quadratic in $P_{\mu ABCD}$ is therefore a standard kinetic term for the physical scalars.
The terms linear in $P_{\mu ABCD}$, including
\begin{align}
&\hat{P}_{\mu ABCD}={P}_{\mu ABCD}+2\sqrt{2}\left(\bar{\psi}^{(L)}_{\mu[A}\lambda^{(R)}_{BCD]}+\frac{1}{24}\epsilon_{ABCDEFGH}\bar{\psi}^{(R)E}_{\mu}\lambda^{(L)FGH}\right)~,
\end{align}
do not contribute to quadratic fluctuations around a background with constant scalars. The covariant derivatives $D_\mu$ that act on fermions are $SU(8)$
covariant so at this point the Lagrangian is manifestly invariant under the local $SU(8)$.

The gauge fields and their duals are
\begin{align}
&G^{MN}_{\mu\nu}=\partial_{\mu}B^{MN}_{\nu}-\partial_{\nu}B^{MN}_{\mu}~,
\label{eqn:GMNdef}
\\
&{\widetilde H}^{\text{(F)}\mu\nu}_{MN}= {4i\over e} {\partial {\cal L}\over\partial G^{MN}_{\mu\nu}}~.
\end{align}
They enter the Lagrangian (\ref{N=8:Lagrangian}) explicitly. Their Pauli couplings are written in terms of
\begin{align}
&\hat{\mathcal{F}}_{AB}=\gamma^{\mu\nu}\mathcal{F}_{AB\mu\nu}~,
\label{eqn:Fab}
\end{align}
where
\begin{align}
& \mathcal{F}_{AB\mu\nu}=\mathcal{F}^{\text{(F)}}_{AB\mu\nu}+\sqrt{2}\left(\bar{\psi}_{[A[\mu}^{(R)}\psi_{[B[\nu}^{(L)}-\frac{1}{\sqrt{2}}\bar{\psi}^{(L)C}_{[\mu}\gamma_{\nu]}\lambda_{ABC}^{(L)}-\frac{1}{288}\epsilon_{ABCDEFGH}\bar{\lambda}^{CDE}_{(L)}\gamma_{\mu\nu}\lambda_{(R)}^{FGH}\right)~,
\label{eqn:FABdef}
\end{align}
with
\begin{align}
\begin{pmatrix}\mathcal{F}^{\text{(F)}}_{AB\mu\nu} \\ \bar{\mathcal{F}}^{\text{(F)}AB}_{\mu\nu}\end{pmatrix}=\frac{1}{\sqrt{2}}\mathcal{V}\begin{pmatrix}G^{MN}_{\mu\nu}+iH^{\text{(F)}}_{MN\mu\nu} \\ G^{MN}_{\mu\nu}-iH^{\text{(F)}}_{MN\mu\nu}\end{pmatrix}~.
\label{eqn:fabGH}
\end{align}
These relatives of the gauge fields encode couplings and $E_{7(7)}$ duality symmetries.
They satisfy the self-duality constraint
\begin{align}
\label{H:constrain}
\mathcal{F}_{\mu\nu AB}=\widetilde{\mathcal{F}}_{\mu\nu AB}~.
\end{align}
This self-duality constraint is a complex equation that relates the real fields $G^{MN}_{\mu\nu}$, $H^{\text{(F)}}_{MN\mu\nu}$ and their duals linearly, with coefficients that depend nonlinearly on scalar fields.
It has a solution of the form
\begin{align}
\label{H:solution}
\widetilde{H}^{\text{(F)}}_{MN\mu\nu}=&-i\left(\mathcal{N}_{MNPQ}G^{-PQ}_{\mu\nu} + {\rm h.c.}\right) +\text{(terms quadratic in fermions)}~,
\end{align}
where the self-dual (anti-self-dual) parts of the field strengths are defined as
\begin{align}
&G^{\pm MN}_{\mu\nu}=\frac{1}{2}\left(G^{MN}_{\mu\nu}\pm\widetilde{G}^{MN}_{\mu\nu}\right)~,
\end{align}
and the gauge coupling function is
\begin{align}
\label{N:definition}
&\mathcal{N}_{MNPQ}=\left(U_{AB}^{\quad MN}-V_{ABMN}\right)^{-1}\left(U_{AB}^{\quad MN}+V_{ABPQ}\right)~.
\end{align}
Using (\ref{H:solution}) for $\widetilde{H}^{\text{(F)}}_{MN\mu\nu}$ and (\ref{eqn:Fab}--\ref{eqn:fabGH}) for $\hat{\mathcal{F}}_{AB}$ we can
eliminate these fields from the Lagrangian (\ref{N=8:Lagrangian}) in favor of the dynamical gauge field $G^{MN}_{\mu\nu}$,
embellished by scalar fields and fermion bilinears.

The relatively complicated classical dynamics of ${\cal N}=8$ SUGRA is due to the interplay between fermion bilinears, duality, and the scalar coset.
These disparate features are all important in our considerations but they largely decouple. For example, although we need the Pauli couplings of fermions,
we need them only for trivial scalars.

In our explicit computations it is convenient to remove the $SU(8)$ gauge redundancy by writing the 56-bein (\ref{eqn:56bein})
in a symmetric gauge
\begin{align}
\mathcal{V}=\exp\begin{pmatrix}0 & W_{ABCD} \\ \bar{W}^{ABCD} & 0\end{pmatrix}~,
\end{align}
where the $70$ complex scalars $W_{ABCD}$ are subject to the constraint
\begin{align}\label{W:constrain}
&\bar{W}^{ABCD}=\frac{1}{24}\epsilon^{ABCDEFGH}W_{EFGH}~.
\end{align}
After fixing the local $SU(8)$ symmetry, the theory still enjoys a global $SU(8)$ symmetry. Moreover, it is linearly realized when compensated by
$SU(8)\subset E_{7(7)}$. We identify this residual global $SU(8)$ as the $R$-symmetry $SU(8)_R$. This identification proves useful repeatedly.
For example, it is according to this residual symmetry that $W_{ABCD}$ transforms as an antisymmetric four-tensor.

 \subsection{The Embedding into ${\cal N}=8$ SUGRA}
 The embedding of the Kaluza-Klein black hole (\ref{eqn:metric}, \ref{eqn:KKh123}, \ref{eqn:KKAKK}) in $\mathcal{N}=8$ SUGRA
 is implemented by
 \begin{align}
 \label{N=8:embedding}
 &\mathring{g}_{\mu\nu}^{(\text{SUGRA})}=g_{\mu\nu}^{(\text{KK})}~,\nonumber\\
 &\mathring{G}^{MN}_{\mu\nu}=\frac{1}{4}\Omega^{MN}F^{(\text{KK})}_{\mu\nu}~,\nonumber\\
 &\mathring{W}_{ABCD}=0~,\nonumber\\
 &(\text{All background fermionic fields})=0~,
 \end{align}
 where
 \begin{align}\arraycolsep=4pt
 \Omega^{MN}=\text{diag}(\epsilon,\epsilon,\epsilon,\epsilon) ~,~~~~ \epsilon=\begin{pmatrix}0&1\\-1&0\end{pmatrix}~.
 \label{eqn:symplectic}
 \end{align}
In this section (and beyond) we shall often declutter formulae by omitting the superscript ``KK" when referring to fields of the seed solution.

To establish the consistency of our embedding, in the following we explicitly check that the $\mathcal{N}=8$ SUGRA equations of motion are
satisfied by the background (\ref{N=8:embedding}). Vanishing fermions satisfy trivially their equations of motion, because they appear at least quadratically in the action.
The equations of motion for the scalars $W_{ABCD}$ take the form
\begin{align}
(\text{Terms at least linear in }\mathring{W}_{ABCD}\text{ or quadratic in fermions}) \nonumber\\
=~3\,\mathring{G}^{+[AB}_{\mu\nu}\mathring{G}^{+CD]\mu\nu}+\frac{1}{8}\epsilon_{ABCDEFGH}\mathring{G}^{-EF}_{\mu\nu}\mathring{G}^{-GH\mu\nu}~.
\end{align}
The scalars $\mathring{W}_{ABCD}$ and the fermions vanish so the right-hand side of the equation must also vanish. Inserting $\mathring{G}^{MN}_{\mu\nu}$ from our embedding (\ref{N=8:embedding}),
we find the condition $F^{(\text{KK})}_{\mu\nu}F^{(\text{KK})\mu\nu}=0$. This condition is satisfied by the seed solution (\ref{EoM:PhiKK2}) because the electric and magnetic charges are equal $P=Q$.
Therefore it is consistent to take all scalars $\mathring{W}_{ABCD}=0$ in $\mathcal{N}=8$ SUGRA.

The $\mathcal{N}=8$ Einstein equation is given by
\begin{eqnarray}\label{EoM:graviton}
R_{\mu\nu}-\frac{1}{2}g_{\mu\nu}R&=&\frac{1}{6}P_{ABCD\{\mu}\bar{P}_{\nu\}}^{ABCD}-\frac{1}{12}g_{\mu\nu}P_{\rho ABCD}\bar{P}^{\rho ABCD}\nonumber\\
&&+\text{Re}(\mathcal{N}_{MNPQ})\left(G^{MN}_{\mu\rho}G_{\nu}^{\ \rho PQ}-\frac{1}{4}g_{\mu\nu}G^{MN}_{\rho\sigma}G^{\rho\sigma PQ}\right)~.
\end{eqnarray}
The vanishing of the scalars $\mathring{W}_{ABCD}=0$ implies
\begin{align}
&\mathring{\mathcal{V}}=\begin{pmatrix}\delta_{[A}^{\ [M}\delta_{B]}^{\ N]} & 0 \\ 0 & \delta^{[A}_{\ [M}\delta^{B]}_{\ N]}\end{pmatrix}~,\qquad
\mathring{\mathcal{N}}_{MNPQ}=\textbf{1}_{MNPQ}~,
\label{eqn:nunsimple}
\end{align}
so the Einstein equation simplifies to
 \begin{align}
 &\mathring{R}_{\mu\nu}-\frac{1}{2}\mathring{g}_{\mu\nu}\mathring{R}=\mathring{G}^{MN}_{\mu\rho}\mathring{G}_{\nu MN}^{\ \rho}-\frac{1}{4}\mathring{g}_{\mu\nu}\mathring{G}^{MN}_{\rho\sigma}\mathring{G}^{\rho\sigma}_{MN}~.
 \end{align}
The embedding (\ref{N=8:embedding}) reduces the right-hand side so that these equations coincide
with the Einstein equation (\ref{EoM:gKK2}) satisfied by the seed solution.

Finally, the equations of motion for the vector fields in ${\cal N}=8$ SUGRA are
\begin{align}\label{EoM:vector}
 D_{\mu}\left(\mathcal{N}_{MNPQ}G^{-\mu\nu PQ}+\mathcal{\bar{N}}_{MNPQ}G^{+\mu\nu PQ}\right)=0~.
\end{align}
The embedding (\ref{N=8:embedding}) and the simplifications (\ref{eqn:nunsimple}) reduce these equations  to the
Maxwell equation $ D_{\mu}F^{(\text{KK})\mu\nu}=0$, consistent with the seed equation of motion (\ref{EoM:AKK2}).

In summary, the equations of motion in $\mathcal{N}=8$ SUGRA are satisfied by the embedding (\ref{N=8:embedding}).
Therefore, for any seed solution that satisfies (\ref{EoM:gKK2}-\ref{EoM:PhiKK2}), the embedding (\ref{N=8:embedding})
gives a solution to $\mathcal{N}=8$ SUGRA. Our primary example is the Kaluza-Klein black hole with dilaton $\Phi^{\text{(KK)}}=0$.

\section{Quadratic Fluctuations in ${\cal N}=8$ SUGRA}\label{sec:fluctuations}
In this section we expand the Lagrangian (\ref{N=8:Lagrangian}) for $\mathcal{N}=8$ SUGRA to quadratic order around the background (\ref{N=8:embedding}).  We reparametrize the fluctuation fields so that they all transform in representations of the global
$USp(8)$ symmetry group preserved by the background. We then partially decouple the quadratic fluctuations into different blocks
corresponding to different representations of $USp(8)$.

\subsection{Global Symmetry of Fluctuations}
\label{sec:reparemeterize}
The $\mathcal{N}=8$ SUGRA theory has a global $SU(8)$ symmetry, as discussed at the end of section \ref{sec:N=8SUGRA}.
The graviton, gravitini, vectors, gaugini, and scalars transform in the representations \textbf{1}, \textbf{8}, \textbf{28}, \textbf{56} and \textbf{70}
of this $SU(8)$ group. The \textbf{28}, \textbf{56}, and \textbf{70}, are realized as antisymmetric combinations of the fundamental
representation \textbf{8}.

A generic background solution does not respect all the symmetries of the theory, so the global $SU(8)$ symmetry is not generally helpful for analyzing fluctuations
around the background. Our embedding (\ref{N=8:embedding}) into ${\cal N}=8$ SUGRA indeed
breaks the $SU(8)$ symmetry  since $\mathring{G}^{MN}_{\mu\nu}=\frac{1}{4}\Omega^{MN}F^{(\text{KK})}_{\mu\nu}$ is not invariant under
the $SU(8)$ group.
However, the matrix $\Omega^{MN}$ (\ref{eqn:symplectic}) can be interpreted as a canonical symplectic form so our embedding respects most of the
global $SU(8)$, it preserves a $USp(8)$ subgroup. Therefore,
different $USp(8)$ representations cannot couple at quadratic order and it greatly simplifies the analysis
to organize fluctuations around the background as representations of $USp(8)$. In the following we analyze one $USp(8)$ representation at a time.

\begin{itemize}
	
\item \emph{Graviton}

The graviton $h_{\mu\nu}=\delta g_{\mu\nu}=g_{\mu\nu}-\mathring{g}_{\mu\nu}$ is a singlet of $SU(8)$ and
remains a singlet of $USp(8)$.

\item \emph{Vectors}

The fluctuations of the gauge fields $\delta G_{\mu\nu}^{MN} = G_{\mu\nu}^{MN} - \mathring{G}^{MN}_{\mu\nu}$ transform in the $\mathbf{28}$ of $SU(8)$ which has the branching rule to $USp(8)$
$\mathbf{28}\rightarrow \mathbf{1}\oplus\mathbf{27}$. We realize this decomposition directly on the fluctuations by defining
\begin{align}
&f_{\mu\nu}=\Omega_{MN}\delta G_{\mu\nu}^{MN}~, \qquad f^{MN}_{\mu\nu}=\delta G_{\mu\nu}^{MN}-\frac{1}{8}\Omega^{MN}f_{\mu\nu}~.
\end{align}
The $f^{MN}_{\mu\nu}$ are $\Omega$-traceless $f^{MN}_{\mu\nu}\Omega_{MN}=0$ by construction so they have only
$2\times(28-1)$ degrees of freedom which transform in the $\mathbf{27}$ of $USp(8)$. The remaining 2 degrees of freedom are in
$f_{\mu\nu}$, which transforms in the $\mathbf{1}$ of $USp(8)$. This decomposition under the global symmetry shows that the graviton can only mix with the ``overall" gauge field $f_{\mu\nu}$ and not  with $f^{MN}_{\mu\nu}$.

\item \emph{Scalars}

The scalars transform in $\mathbf{70}$ of $SU(8)$ and the branching rule to $USp(8)$ is $\mathbf{70}\rightarrow \mathbf{1}\oplus\mathbf{27}\oplus\mathbf{42}$. We realize this decomposition by defining
\begin{align}
&W'=W_{ABCD}\Omega^{AB}\Omega^{CD}~, \quad W'_{AB}=W_{ABCD}\Omega^{CD}-\frac{1}{8}W'\Omega_{AB}~,\nonumber\\
&W'_{ABCD}=W_{ABCD}-\frac{3}{2}W'_{[AB}\Omega_{CD]}-\frac{1}{16}W'\Omega_{[AB}\Omega_{CD]}~.
\end{align}
$W'_{ABCD}$ is antisymmetric in all indices and $\Omega$-traceless on any pair or pairs, so it is in the
$\mathbf{42}$ of $USp(8)$. $W'_{AB}$ is antisymmetric, $\Omega$-traceless, and hence in the $\mathbf{27}$ of $USp(8)$. The
remainder $W'$ has no index and is in the ${\bf 1}$ of $USp(8)$.
The obvious construction of an antisymmetric four-tensor representation of $SU(8)$ has $70$ {\it complex} degrees of freedom, but
the scalars $W_{ABCD}$ in ${\cal N}=8$ SUGRA have $70$ {\it real} degrees of freedom that realize an irreducible representation,
as implemented by the reality constraint (\ref{W:constrain}). The decomposition of this reality constraint under $SU(8)\to USp(8)$
shows that the scalar $W'$ that couples to gravity is real $\overline{W}'=W'$, as expected
from Kaluza-Klein theory. It also implies the reality condition on the four-tensor
\begin{equation}\label{eqn:wabcdreality}
{\overline W}'^{ABCD}=\frac{1}{24}\epsilon^{ABCDEFGH}W'_{EFGH}~,
\end{equation}
and an analogous condition on the two-tensor $W^{\prime AB}$. An interesting aspect of these reality conditions is that, just like the KK block must
couple to a scalar (as opposed to a pseudoscalar), the condition on the $USp(8)$ four-tensor demonstrates that the scalar moduli must comprise exactly $22$ scalars and $20$ pseudoscalars. The vector multiplet couples vectors and scalars/pseudoscalars precisely so that it restores
the overall balance between scalars and pseudoscalars required by ${\cal N}=8$ SUGRA, with $12$ scalars and $15$ pseudoscalars.

The distinctions between scalars and pseudoscalars are interesting because these details must be reproduced by viable microscopic models of black holes.
Extrapolations far off extremality of phenomenological models that are motivated by the BPS limit lead to entropy formulae
\cite{Obers:1998fb, Larsen:1999dh, Cvetic:2014sxa}
with moduli dependence that is very similar but not identical to the result found here. It would be interesting to construct a model for non-extremal black holes that combines the features of the BPS and the non-BPS branch.

\item \emph{Gravitini}

The gravitini $\psi_{A\mu}$ transform in the fundamental $\mathbf{8}$ of $SU(8)$. The gravitini only carry one $SU(8)$ index which cannot be contracted with the symplectic form $\Omega^{AB}$. Therefore, the gravitini also transform in the $\mathbf{8}$ of $USp(8)$.

\item \emph{Gaugini}

The gaugini $\lambda_{ABC}$ of ${\cal N}=8$ SUGRA transform in the $\mathbf{56}$ of the global $SU(8)$. The branching rule to  $USp(8)$ is
$\mathbf{56}\rightarrow \mathbf{8}\oplus\mathbf{48}$.
We can realize this decomposition by introducing
\begin{align}
\lambda'_A &= {1\over\sqrt{12}} \lambda_{ABC}\Omega^{BC}~,
\label{eqn:lamAdef}
\end{align}
and
\begin{align}
\lambda'_{ABC} &=\lambda_{ABC}-\frac{1}{8}(\lambda_{ADE}\Omega^{DE})\Omega_{BC}~.
\label{eqn:lamABCdef}
\end{align}
The gaugini $\lambda'_A$ transform in the ${\bf 8}$ of $USp(8)$. We will find that these gaugini are coupled to the gravitini. This is allowed because they have the same quantum numbers under the global $USp(8)$. The normalization $1/\sqrt{12}$ introduced in (\ref{eqn:lamAdef}) ensures that the gaugini
retain a canonical kinetic term after the field redefinition.

The gaugini $\lambda'_{ABC}$ introduced in (\ref{eqn:lamABCdef}) satisfy the constraint $\lambda'_{ABC}\Omega^{BC}=0$. This ensures that they
transform in the ${\bf 48}$ of $USp(8)$. No other fields transform in the same way under the global symmetry so these gaugini decouple from other fields.
They can of course mix among themselves and we will find that they do in fact have
nontrivial Pauli couplings. However, the normalization of the fields is inconsequential
and we have retained the normalization inherited from the full ${\cal N}=8$ SUGRA.
\end{itemize}

Table \ref{Rep of USp} summarizes the decomposition of quadratic fluctuations according to their representations under the global $USp(8)$
that is preserved by the background.
\medskip

\renewcommand\arraystretch{1.5}
\begin{table}[h]
\centering
\begin{tabular}{|c|p{3cm}<{\centering}|}
\hline
Representations & Fields \\ \hline\hline\textbf{1}
& $h_{\mu\nu},~ f_{\mu\nu},~ W'$\\ \hline
\textbf{8} & $\psi_{A\mu},~ \lambda'_A$ \\ \hline
\textbf{27} & $f^{\mu\nu}_{AB},~ W'_{AB}$ \\ \hline
\textbf{42} & $W'_{ABCD}$ \\ \hline
\textbf{48} & $\lambda'_{ABC}$ \\
\hline
\end{tabular}
\caption{The $USp(8)$ representation content of the quadratic fluctuations.}
\label{Rep of USp}
\end{table}
%

\subsection{The Decoupled Fluctuations}
\label{sec:decoupledfluc}

The quadratic fluctuations around any bosonic background decouple into a bosonic part $\delta^{2}\mathcal{L}_{\text{bosons}}$
and a fermionic part $\delta^{2}\mathcal{L}_{\text{fermions}}$ because fermions always appear quadratically in the Lagrangian.
As we expand the Lagrangian (\ref{N=8:Lagrangian}) around the background (\ref{N=8:embedding}) to quadratic order, these parts
further decouple into representations of the preserved $USp(8)$ global symmetry.

\medskip

The bosonic fluctuations therefore decouple into three blocks
\begin{align}
&\delta^{2}\mathcal{L}^{(\mathcal{N}=8)}_{\text{bosons}}=\delta^{2}\mathcal{L}^{(\mathcal{N}=8)}_{\text{KK}}+\delta^{2}\mathcal{L}^{(\mathcal{N}=8)}_{\text{vector}}+\delta^{2}\mathcal{L}^{(\mathcal{N}=8)}_{\text{scalar}}~.
\end{align}

\begin{itemize}
	
	\item \emph{KK block}
	
The first block $\delta^{2}\mathcal{L}^{(\mathcal{N}=8)}_{\text{KK}}$, which we call the ``KK block",
consists of all fields that are singlets of $USp(8)$: the graviton $h_{\mu\nu}$, $1$ vector with field strength $f_{\mu\nu}$, and $1$ scalar $W'$.
The Lagrangian for this block is given by
\begin{eqnarray}
\label{N=8:CMb}
e^{-1}\delta^{2}\mathcal{L}^{(\mathcal{N}=8)}_{\text{KK}}&=&\bar{h}^{\mu\nu}\Box\bar{h}_{\mu\nu}-\frac{1}{4}h\Box h+2\bar{h}^{\mu\nu}\bar{h}^{\rho\sigma}R_{\mu\rho\nu\sigma}-2\bar{h}^{\mu\nu}\bar{h}_{\mu\rho}R^{\rho}_{\ \nu}-h\bar{h}^{\mu\nu}R_{\mu\nu}\nonumber\\
&&-F^{}_{\mu\nu}F^{}_{\rho\sigma}\bar{h}^{\mu\rho}\bar{h}^{\nu\sigma}+a^{\mu}\left(\Box g_{\mu\nu}-R_{\mu\nu}\right)a^{\nu}+2\sqrt{2}F^{\ \rho}_{\nu}f_{\mu\rho}\bar{h}^{\mu\nu}\nonumber\\
&&-4\partial_{\mu}\phi\partial^{\mu}\phi+2\sqrt{3}F^{\mu\nu}f_{\mu\nu}\phi-4\sqrt{6}R_{\mu\nu}\bar{h}^{\mu\nu}\phi~,
\end{eqnarray}
after the fields were redefined as $h_{\mu\nu}\rightarrow {\sqrt{2}}h_{\mu\nu}$, $f_{\mu\nu}\rightarrow {4}f_{\mu\nu}$, and $\phi=-\frac{1}{8\sqrt{3}}W'$.
We also decomposed the graviton into its trace $h=g^{\rho\sigma}h_{\rho\sigma}$ and its traceless part $\bar{h}_{\mu\nu}={h}_{\mu\nu}-\frac{1}{4}g_{\mu\nu}g^{\rho\sigma}h_{\rho\sigma}$, and further included the gauge-fixing term
\begin{eqnarray}
e^{-1}\mathcal{L}_{\text{g.f.}}&=&-\left(D^{\mu}\bar{h}_{\mu\rho}-\frac{1}{2}D_{\rho}h\right)\left(D^{\nu}\bar{h}_{\nu}^{\ \rho}-\frac{1}{2}D^{\rho}h\right)-\left(D^{\mu}a_{\mu}\right)^{2}~.
\end{eqnarray}
The rather complicated Lagrangian (\ref{N=8:CMb}) represents the theory of fluctuations around any solution of
Kaluza-Klein theory (\ref{eqn:KKtheory}) with constant dilaton. The fields $f_{\mu\nu}$ and $\phi$ correspond
to the fluctuations of the field strength and the dilaton. The gauge-fixed theory (\ref{N=8:CMb}) must be completed with additional ghost terms. We discuss those in Appendix \ref{app:a4}.

\item \emph{Vector blocks}

The second block $\delta^{2}\mathcal{L}^{(\mathcal{N}=8)}_{\text{vector}}$ consists of all fields that transform in the $\mathbf{27}$ of $USp(8)$: $f^{\mu\nu}_{AB}$ and $W'_{AB}$. We use  $f^{\mu\nu}_{a}$  and  $W'_{a}$ to denote the 27 independent vectors and scalars respectively. It includes two slightly different parts. One part has $12$ copies of a vector coupled to a scalar $W'^{(\text{R})}_{a}$ with the Lagrangian
\begin{align}\label{N=8:vector}
&e^{-1}\delta^{2}\mathcal{L}^{(\mathcal{N}=8)(\text{R})}_{\text{vector}}=-\frac{1}{2}\partial^{\mu}W'^{(\text{R})}_{a}\partial_{\mu}W'^{(\text{R})}_{a}-f^{\mu\nu}_{a}f_{a\mu\nu}-W'^{(\text{R})}_{a}f_{a\mu\nu}F^{\mu\nu}~, ~ a=1,...,12~,
\end{align}
and the other has $15$ copies of a vector coupled to a pseudoscalar $W'^{(\text{P})}_{a}$ given by
\begin{align}
\label{N=8:Pvector}
&e^{-1}\delta^{2}\mathcal{L}^{(\mathcal{N}=8)(\text{P})}_{\text{vector}}=-\frac{1}{2}\partial^{\mu}W'^{(\text{P})}_{a}\partial_{\mu}W'^{(\text{P})}_{a}-f^{\mu\nu}_{a}f_{a\mu\nu}-iW'^{(\text{P})}_{a}f_{a\mu\nu}\widetilde{F}^{\mu\nu}~, ~ a=13,...,27.
\end{align}
Although these two Lagrangians are distinct, they give equations of motion that are equivalent under a duality transformation. This is consistent with the fact
that $SU(8)$ duality symmetry is the diagonal combination of local $SU(8)$ and global $E_{7(7)}$ duality symmetry, where the latter is not realized at the level of the Lagrangian.

\item \emph{Scalar blocks}

The last bosonic block $\delta^{2}\mathcal{L}^{(\mathcal{N}=8)}_{\text{scalar}}$ consists of the remaining $42$ scalars, transforming in the $\mathbf{42}$ of $USp(8)$. There are no other bosonic fields with the same quantum numbers so, these fields can only couple to themselves. The explicit expansion around the
background (\ref{EoM:gKK2}-\ref{EoM:PhiKK2}) shows that all these scalars are in fact minimally coupled
\begin{align}\label{N=8:scalar}
e^{-1}\delta^{2}\mathcal{L}^{(\mathcal{N}=8)}_{\text{scalar}}=-\frac{1}{24}\partial^{\mu}W'_{ABCD}\partial_{\mu}\overline{W}'^{ABCD}~.
\end{align}
\end{itemize}

We now turn to the quadratic fluctuations for the fermions. Since they appear at least quadratically in the Lagrangian the bosonic fields can be fixed to their
background values. In this case, the ${\cal N}=8$ SUGRA Lagrangian (\ref{N=8:Lagrangian}) simplifies to
\begin{eqnarray}\label{N=8:FermionicLagrangian}
e^{-1}\delta^{2}\mathcal{L}^{(\mathcal{N}=8)}_{\text{fermions}}&=&-\frac{1}{2}\bar{\psi}_{A\mu}\gamma^{\mu\nu\rho}D_{\nu}\psi_{A\rho}-\frac{1}{12}\bar{\lambda}_{ABC}\gamma^{\mu}D_{\mu}\lambda_{ABC}+\frac{1}{4\sqrt{2}}\bar{\psi}_{A\mu}\gamma^{\nu}\mathring{\mathcal{F}}_{AB}\gamma^{\mu}\psi_{\nu B}\nonumber\\
&&-\frac{1}{8}\bar{\psi}_{C\mu}\mathring{\mathcal{F}}_{AB}\gamma^{\mu}\lambda_{ABC}+\frac{1}{288\sqrt{2}}\epsilon^{ABCDEFGH}\bar{\lambda}_{ABC}\mathring{\mathcal{F}}_{DE}\lambda_{FGH}~,
\end{eqnarray}
where all fermions are in Majorana form and
\begin{eqnarray}
\mathring{\mathcal{F}}_{AB}&=&\frac{1}{\sqrt{2}}\left(\mathring{G}_{AB\mu\nu}+\gamma_{5}\mathring{\widetilde{G}}_{AB\mu\nu}\right)\gamma^{\mu\nu}=\frac{1}{2\sqrt{2}}\Omega_{AB}F_{\mu\nu}\gamma^{\mu\nu}~.
\end{eqnarray}
The field redefinitions introduced in section \ref{sec:reparemeterize} decouple this Lagrangian as
\begin{equation}
\delta^{2}\mathcal{L}^{(\mathcal{N}=8)}_{\text{fermions}}=\delta^{2}\mathcal{L}^{(\mathcal{N}=8)}_{\text{gravitino}}+\delta^{2}\mathcal{L}^{(\mathcal{N}=8)}_{\text{gaugino}}~.
\end{equation}

\begin{itemize}
	\item \emph{Gravitino blocks}
	
The first block $\delta^{2}\mathcal{L}^{(\mathcal{N}=8)}_{\text{gravitino}}$ consists of the $8$ gravitini $\psi_{A\mu}$ and the $8$
gaugini $\lambda'_A$ singled out by the projection (\ref{eqn:lamAdef}). The gravitini and the gaugini both transform
in $\mathbf{8}$ of $USp(8)$ and couple through the Lagrangian
\begin{eqnarray}\label{N=8:gravitino}
e^{-1}\delta^{2}\mathcal{L}^{(\mathcal{N}=8)}_{\text{gravitino}}&=&-\bar{\psi}_{A\mu}\gamma^{\mu\nu\rho}D_{\nu}\psi_{A\rho}-\bar{\lambda}'_A\gamma^{\mu}D_{\mu}\lambda'_A+\frac{1}{4}\Omega^{AB}\bar{\psi}_{A\mu}\left(F^{\mu\nu}+\gamma_{5}\widetilde{F}^{\mu\nu}\right)\psi_{B\nu}\nonumber\\
&&-\frac{\sqrt{6}}{8}\bar{\psi}_{A\mu}F_{\rho\sigma}\gamma^{\rho\sigma}\gamma^{\mu}\lambda'_A+\frac{1}{4}\Omega^{AB}\bar{\lambda}'_A F_{\rho\sigma}\gamma^{\rho\sigma}\lambda'_B~.
\end{eqnarray}
The indices take values $A, B= 1,\ldots 8$. However, this block actually decouples into $4$ identical pairs, with a single pair comprising
two gravitini and two gaugini. The canonical pair is identified by restricting the indices to $A, B= 1,2$ and
so $\Omega_{AB}\to\epsilon_{AB}$. The other pairs correspond to $A, B= 3,4$, $A, B= 5,6$, and $A, B= 7,8$.

\item \emph{Gaugino blocks}

The second block $\delta^{2}\mathcal{L}^{(\mathcal{N}=8)}_{\text{gaugino}}$ consists of the $48$ gaugini (\ref{eqn:lamABCdef}) that transform in the
$\mathbf{48}$ of $USp(8)$. These $48$ gaugini decompose into $24$ identical groups that decouple from one another.
Each group has $2$ gaugini and a Lagrangian
given by
\begin{align}\label{N=8:gaugino}
e^{-1}\delta^{2}\mathcal{L}^{(\mathcal{N}=8)}_{\text{gaugino}}=&-\bar{\lambda}_{a}\gamma^{\mu}D_{\mu}\lambda_{a}-\frac{1}{8}\epsilon^{ab}\bar{\lambda}_{a}F_{\mu\nu}\gamma^{\mu\nu}\lambda_{b}~,
\end{align}
where $a,b=1,2$ denote the 2 different gaugini in one group. It is interesting that {\it no} fermions in the theory are minimally coupled. Moreover, the
numerical strength of the Pauli couplings to black holes on the non-BPS branch are different from the corresponding Pauli couplings for fermions on the BPS branch \cite{Charles:2015eha}.

\end{itemize}

\subsection{Summary of Quadratic Fluctuations}\label{N=8:Summary}

In the previous sections we defined a seed solution (\ref{EoM:gKK2}-\ref{EoM:PhiKK2}) of Kaluza-Klein theory
with vanishing dilaton and embedded it into $\mathcal{N}=8$ SUGRA through (\ref{N=8:embedding}). In this section, we have studied fluctuations around the background by expanding the ${\cal N}=8$ SUGRA Lagrangian (\ref{N=8:Lagrangian})
to quadratic order. In section \ref{sec:reparemeterize}, we decomposed the fluctuations in representations of
the $USp(8)$ symmetry preserved by the background. In section \ref{sec:decoupledfluc}, we have decoupled the quadratic fluctuations into blocks corresponding to distinct representations of $USp(8)$. They are summarized in Table \ref{FluctTable}.

\begin{table}[H]
	\centering
	\begin{tabular}{|c|c|c|c|c|c|}
		\hline
		Degeneracy &	Multiplet 	& Block content & $USp(8)$  & Lagrangian \\\hline\hline
		1 & KK block & 1 graviton, 1 vector, 1 scalar& {\bf 1} & (\ref{N=8:CMb}) \\\hline
		27 &	Vector block  & 1 vector and 1 (pseudo)scalar&  {\bf 27} & (\ref{N=8:vector})  \\\hline
		42 & Scalar block & 1 real scalar&  {\bf 42} & (\ref{N=8:scalar})  \\\hline
		4 &	Gravitino block &2 gravitini and 2 gaugini &  {\bf 8} & (\ref{N=8:gravitino})  \\\hline
		24 & 	Gaugino block &2 gaugini& {\bf 48} & (\ref{N=8:gaugino}) \\
		\hline
	\end{tabular}
	
	\caption{ Decoupled quadratic fluctuations in $\cN=8$ supergravity around the KK black hole.}\label{FluctTable}
\end{table}

\section{Consistent Truncations of ${\cal N}=8$ SUGRA}\label{sec:n8}

In this section we present consistent truncations from ${\cal N}=8$ SUGRA to ${\cal N}=6$, ${\cal N}=4$, ${\cal N}=2$ and
${\cal N}=0$. These truncations are well adapted to the KK black hole in that all its nontrivial fields are retained.
In other words, the truncations amount to removal of fields that are trivial in the background solution.

It is easy to analyse the spectrum of quadratic fluctuations around the KK black hole in the truncated theories. In each case some of the fluctuating fields
are removed, but always consistently so that blocks of fields that couple to each other are either all retained or all removed. Therefore,
the fluctuation spectrum in all these theories can be described in terms of the same simple blocks that appear in ${\cal N}=8$ supergravity.
For these truncations the entire dependence on the theory is encoded in the degeneracy of each type of block.
They are summarized in Table \ref{tab:degeneracy}.

\medskip
\begin{table}[h]
\centering
\begin{tabular}{|c||c|c|c|c|c|}
\hline
Multiplet $\backslash$ Theory & ${\cal N}=8$ & ${\cal N}=6$  & ${\cal N}=4$ & ${\cal N}=2$ & ${\cal N}=0$ \cr
\hline
\hline
{\rm KK block} & $1$ &  $1$ & $1$ & $1$ & $1$\cr
\hline
{\rm Gravitino block} & $4$ & $3$ & $2$ & $1$ & $0$ \cr
\hline
{\rm Vector block}  & $27$ & $15$ & $n+5$ & $n_V$ & $0$ \cr
\hline
{\rm Gaugino block} &  $24$ & $10$  & $2n$ & $n_V-1$ & $0$ \cr
\hline
{\rm Scalar block } & $42$  & $14$  & $5n-4$ & $n_V-1$ & $0$ \cr
\hline
\end{tabular}
\caption{The degeneracy of multiplets in the spectrum of quadratic fluctuations around the KK black hole embedded in various theories.
For ${\cal N}=4$, the integer $n$ is the number of ${\cal N}=4$ matter multiplets. For ${\cal N}=2$, the integer $n_V$
refers to the $ST(n_V-1)$ model.}
\label{tab:degeneracy}
\end{table}

All the truncations in this section heavily utilize the $SU(8)_R$ global symmetry of ${\cal N}=8$ supergravity. We therefore
recall from the outset that the gravitons, gravitini, vectors, gaugini, and scalars transform in the irreducible representations
${\bf 1}$, ${\bf 8}$, ${\bf 28}$, ${\bf 56}$, ${\bf 70}$ of $SU(8)_R$.

\subsection{The ${\cal N}=6$ Truncation}

The ${\cal N}=6$ truncation restricts ${\cal N}=8$ SUGRA to fields that are {\it even} under the $SU(8)_R$ element ${\rm diag} (I_6,-I_2)$.
This projection preserves ${\cal N}=6$ local supersymmetry since the $8$ gravitini of ${\cal N}=8$ SUGRA are in
the fundamental ${\bf 8}$ of $SU(8)_R$ and so exactly two gravitini are odd under ${\rm diag} (I_6,-I_2)$ and projected out.
The branching rules of the matter multiplets under $SU(8)_R\to SU(6)_R\times SU(2)_{\rm matter}$ are
\begin{eqnarray}
{\bf 70} &\to & ({\bf 15},{\bf 1})\oplus ({\bf {\overline{15}}},{\bf 1})\oplus ({\bf 20},{\bf 2})~, \cr
{\bf 56} &\to & ({\bf 20},{\bf 1})\oplus ({\bf 15},{\bf 2})\oplus({\bf 6}, {\bf 1}) ~, \cr
{\bf 28} &\to & ({\bf 15},{\bf 1})\oplus ({\bf 6},{\bf 2}) \oplus ({\bf 1},{\bf 1})~.
\label{eqn:truncation}
\end{eqnarray}
These branching rules follow from decomposition of the $SU(8)_R$ four-tensor $T_{ABCD}$ (${\bf 70}$), the three-tensor $T_{ABC}$ (${\bf 56}$),
and the two-tensor $T_{AB}$ (${\bf 28}$), by splitting the $SU(8)_R$ indices as $A,B,...\to (\alpha,a),(\beta,b),...$ where the lower case indices
refer to $SU(2)_{\rm matter}$ (greek) and $SU(6)_R$   (latin). The truncation to ${\cal N}=6$ SUGRA retains only the fields that are invariant
under $SU(2)_{\rm matter}$ so fields in the ${\bf 2}$ are removed.
Therefore the truncated theory has $30$ scalar fields, $26$ gaugini, and $16$ vector fields. Taking the $6$ gravitini and the graviton
into account as well, the total field content comprises $64$ bosonic and $64$ fermionic degrees of freedom.

The claim that the truncation is {\it consistent} means that the equations of motion {\it of the retained fields} are sufficient to guarantee that {\it all} equations of motion are satisfied, as long as the removed fields vanish. In general, the primary obstacle to truncation is that the equations of motion for the
omitted fields may fail. This is addressed here because the equations of motion for fields in the
${\bf 2}$ of $SU(2)_{\rm matter}$ only involve terms in the ${\bf 2}$. Therefore their equations of motion are satisfied when all fields
in the ${\bf 2}$ vanish.

Our interest in the consistent truncation of ${\cal N}=8$ SUGRA to ${\cal N}=6$ SUGRA is the application to the KK black hole.
The embedding (\ref{N=8:embedding}) of the Kaluza-Klein black hole into ${\cal N}=8$ SUGRA turns on the four field strengths on the
skew-diagonal of the ${\bf 28}$ (which is realized by an antisymmetric $8\times 8$ matrix of field strengths $F_{AB}$).
The entries on the skew diagonal are all contained in the $SU(6)_R\times SU(2)_{\rm matter}$ subgroup of $SU(8)_R$, because the antisymmetric representation of $SU(2)$ is trivial. The embedding of the KK black hole in ${\cal N}=8$ SUGRA therefore defines an embedding
in ${\cal N}=6$ SUGRA as well. In other words, the truncation and the embedding are {\it compatible}.

We can find the spectrum of quadratic fluctuations in ${\cal N}=6$ SUGRA either by truncating the spectrum determined in
the ${\cal N}=8$ SUGRA context, or by directly analyzing the spectrum of fluctuations around the ${\cal N}=6$ solution.
Consistency demands that these procedures agree.

We begin from the $SU(6)$ content of ${\cal N}=6$ SUGRA: ${\bf 1}$ graviton, ${\bf 6}$ gravitini, ${\bf15}\oplus{\bf 1}$
vectors, ${\bf 20}\oplus{\bf 6}$ gaugini, and $2({\bf 15})$ scalars. The KK black hole in ${\cal N}=6$ SUGRA
breaks the global symmetry $SU(6)\to USp(6)$. Therefore, the quadratic fluctuations around the background need not respect the $SU(6)$
symmetry, but they must respect the $USp(6)$. Their $USp(6)$ content is: ${\bf 1}$ graviton, ${\bf 6}$ gravitini, ${\bf14}\oplus 2({\bf 1})$ vectors, ${\bf 14}\oplus 2({\bf 6})$ gaugini, $2({\bf 14}\oplus {\bf 1})$ scalars.
The black hole background breaks Lorentz invariance so the equations of motion for fluctuations generally mix Lorentz representations, as we have seen explicitly in section \ref{sec:fluctuations},
but they always preserve global symmetries. In the present context the mixing combines the fields into ${\bf 1}$ KK block
(gravity + 1 vector + 1 scalar), 3 gravitino blocks (1 gravitino + 1 gaugino) (transforming in the ${\bf 6}$), ${\bf 14}\oplus {\bf 1}$ vector blocks (1 vector + 1 scalar),
$10$ gaugino blocks (transforming in the ${\bf 14}\oplus {\bf 6}$), and ${\bf 14}$ (minimally coupled) scalars.

To verify these claims and find the specific couplings for each block, we could analyze the equations of motion for ${\cal N}=6$
SUGRA using the methods of section \ref{sec:fluctuations}. However, no new computations are needed because it is clear that the fields in the
truncated theory are a subset of those in ${\cal N}=8$ SUGRA. In that context we established that the fluctuations decompose into ${\bf 1}$ (KK block),
${\bf 8}$ (gravitini mixing with gaugini), ${\bf 27}$ (vectors mixing with scalars), ${\bf 24}$ (gaugini with Pauli couplings to the background),
and ${\bf 42}$ (minimal scalars) of the $USp(8)$ that is preserved by the background. The consistent truncation to
${\cal N}=6$ SUGRA removes some of these fluctuations as it projects the global symmetry $USp(8)\to USp(6)$. This rule not only
establishes the mixing claimed in the preceding paragraph but also shows that all couplings must be the same in the ${\cal N}=8$
and ${\cal N}=6$ theories. It is only the degeneracy of each type of block that is reduced by the truncation.

\subsection{The ${\cal N}=4$ Truncation}
\label{sec:N=4sec}

The ${\cal N}=4$ truncation restricts ${\cal N}=8$ SUGRA to fields that are even under the $SU(8)_R$ element ${\rm diag} (I_4,-I_4)$.
This projection breaks the global symmetry $SU(8)_R\to SU(4)_R\times SU(4)_{\rm matter}$. It preserves ${\cal N}=4$ local supersymmetry since
the $8$ gravitini of ${\cal N}=8$ SUGRA are in the ${\bf 4}$ of $SU(4)_R$. The branching rules of the matter multiplets under the symmetry breaking are
\begin{eqnarray}
{\bf 70} &\to & 2({\bf 1},{\bf 1})\oplus ({\bf 6},{\bf 6})\oplus ({\bf 4},{\bf {\bar 4}})\oplus ({\bf {\bar 4}}, {\bf 4}) ~, \cr
{\bf 56} &\to & ({\bf {\bar 4}},{\bf 1})\oplus ({\bf 6},{\bf 4})\oplus ({\bf 4}, {\bf 6}) \oplus ({\bf 1}, {\bf {\bar 4}}) ~, \cr
{\bf 28} &\to & ({\bf 1},{\bf 6})\oplus ({\bf 6},{\bf 1})\oplus ({\bf 4},{\bf 4}) ~.
\label{eqn:bosetruncation}
\end{eqnarray}
The consistent truncation preserving ${\cal N}=4$ supersymmetry is defined by omission of all fields in the ${\bf 4}$ (or ${\bar {\bf 4}}$) of
$SU(4)_{\rm matter}$.

There is a unique supergravity with $n$ ${\cal N}=4$ matter multiplets. It has a global $SU(4)_R$ symmetry that acts on its supercharges and also a global $SO(n)_{\rm matter}$ that reflects the equivalence of all matter multiplets. The consistent truncation of ${\cal N}=8$ by the element
${\rm diag} (I_4,-I_4)$ retains a $SU(4)_R\times SU(4)_{\rm matter}$ symmetry so, recalling that $SO(6)$ and $SU(4)$ are equivalent as
Lie algebras, the truncated theory must be ${\cal N}=4$ SUGRA with $n=6$ matter multiplets.

Several important features of ${\cal N}=4$ SUGRA are succinctly summarized by the scalar coset
\begin{equation}
\frac{SU(1,1)}{U(1)}\times \frac{SO(6,n)}{SO(6)\times SO(n)}~.
\label{eqn:scalarcoset}
\end{equation}
It has dimension $6n+2$ with scalars transforming in $2({\bf 1},{\bf 1})\oplus ({\bf 6},{\bf n})$ under $SU(4)_R \times SO(n)_{\rm matter}$. It also encodes the
$SU(1,1)\simeq SL(2)$ electromagnetic duality of the $6+n$ vector fields in the fundamental of $SO(6,n)$. The representation content
obtained by removal of ${\bf 4}$ (and ${\bar {\bf 4}}$) from the branchings (\ref{eqn:bosetruncation}) is consistent with these expectations when $n=6$.

The ${\cal N}=4$ truncation has a natural interpretation in perturbative Type II string theory. There is a simple duality frame where the diagonal
element ${\rm diag} (I_4,-I_4)$ changes the sign on the ${\rm RR}$ sector and interchanges the ${\rm RNS}$ and ${\rm NSR}$ sectors;
so the consistent truncation projects on to the common sector of Type IIA and Type IIB supergravity. The complete string theory
orbifold includes twisted sectors as well. It is conveniently implemented by a flip of the GSO projection and is equivalent to T-duality
between Type IIA and Type IIB string theory.

The embedding of the KK black hole into ${\cal N}=8$ SUGRA is compatible with the truncation to ${\cal N}=4$ SUGRA:
the four field strengths on the skew-diagonal of the ${\bf 28}$ are all contained in the $SU(4)_R\times SU(4)_{\rm matter}$
subgroup of $SU(8)_R$ and therefore retained in the truncation to ${\cal N}=4$ SUGRA. The embedding
of the KK black hole in ${\cal N}=8$ SUGRA therefore defines an embedding in  ${\cal N}=4$ SUGRA as well. The consistent truncation
just removes fields that are not excited by the KK black hole in ${\cal N}=8$ SUGRA.

The quadratic fluctuations around the KK black hole in ${\cal N}=8$ SUGRA similarly project on to the ${\cal N}=4$ setting. As discussed
in section \ref{sec:fluctuations},
the KK black hole in ${\cal N}=8$ SUGRA breaks the global symmetry $SU(8)_R\to USp(8)$ and this symmetry breaking pattern greatly constrains
the spectrum of fluctuations around the black hole. Moreover, the symmetry breaking pattern
is largely preserved by the consistent truncation: the analogous breaking pattern in ${\cal N}=4$ SUGRA is
$SU(4)_R\times SU(4)_{\rm matter}\to USp(4)_R\times USp(4)_{\rm matter}$. For example, the entire KK block (with a graviton, a vector, and a scalar), identified as the ${\bf 1}$ of $USp(8)$, is unchanged by the consistent truncation.

The $27$ vector blocks (\ref{N=8:vector}-\ref{N=8:Pvector}), each with a vector coupled to a scalar, are perturbations of the $8\times 8$ matrix of field strengths $F_{AB}$ after its
symplectic trace is removed. The branching (\ref{eqn:bosetruncation}) of the ${\bf 28}$ under $SU(4)_R\times SU(4)_{\rm matter}$ shows that $16$ vector blocks are
projected out by the truncation. None of these are affected by the symplectic trace so $27-16=11$ vector blocks remain in
${\cal N}=4$ SUGRA. Among the $38$ scalars from the coset (\ref{eqn:scalarcoset}) with $n=6$ there is $1$ coupled to gravity and $11$ that couple to the
vectors, so $26$ minimally coupled scalars remain. They parametrize the coset
\begin{equation}
SU(1,1)\times \frac{SO(5,5)}{USp(4)\times USp(4)}~.
\end{equation}
The fermionic sector is simpler because the truncation removes exactly one half of the fermions. The retained fermions are essentially identical to those
that are projected away, they differ at most in their chirality and the KK black holes is insensitive to this distinction. The quadratic fluctuations for the fermions in ${\cal N}=8$ SUGRA are $4$ gravitino pairs
(with each pair including two gravitini coupled to two Weyl fermions, a total of 32 degrees of freedom) and $24$ gaugino pairs with
Pauli couplings to the background field strength. In ${\cal N}=4$ SUGRA with $6$ matter multiplets
there are $4$ gravitino pairs and $12$ gaugino pairs.

There is a simple extension of these results to the case of ${\cal N}=4$ SUGRA with $n\neq 6$ matter multiplets. For this generalization,
we recast the symmetry breaking by the field strengths that have been designated ${\cal N}=4$ matter as
$SO(6)_{\rm matter}\to SO(5)_{\rm matter}$ using the equivalences $SU(4)=SO(6)$ and $USp(4)=SO(5)$ as Lie algebras.
In this form the symmetry breaking just amounts to picking the direction of a vector on an $S^{5}$.
We can equally consider any number $n$ of matter fields and break the symmetry $SO(n)_{\rm matter}\to SO(n-1)_{\rm matter}$
by picking a vector on $S^{n-1}$. The only restriction is $n\geq 1$ in order to ensure that there is a direction to pick in the
first place. This more general construction gives the scalar manifold
\begin{equation}
SU(1,1)\times \frac{SO(5,n-1)}{SO(5)\times SO(n-1)}~.
\label{eqn:nonBPScoset}
\end{equation}
In particular, it has $5n-4$ dimensions, each corresponding to a minimally coupled scalar field. The duality group read off from the numerator correctly indicates $n+5$ vector fields, not counting the one coupling to gravity. Each of these vector fields
couples to a scalar field, as in (\ref{N=8:vector}-\ref{N=8:Pvector}).

The black hole attractor mechanism offers a perspective on the scalar coset (\ref{eqn:nonBPScoset}). The attractor mechanism is usually
formulated in the context of extremal black holes in ${\cal N}\geq 2$ supergravity where it determines the value of some of the scalars at the horizon in terms of
black hole charges. Importantly, the attractor mechanism generally leaves other scalars undetermined. Such undetermined scalars can take any value, so they are moduli. The hyper-scalars in ${\cal N}=2$ BPS black hole backgrounds are well-known examples of black hole moduli.

In the case of extremal (but non-supersymmetric) black holes in ${\cal N}\geq 2$ supergravity the moduli space is determined by the
centralizer remaining after extremization of the black hole potential over the full moduli space of the theory. The result for non-BPS
black holes in ${\cal N}=4$ supergravity was obtained in  \cite{Andrianopoli:2006ub} and agrees with (\ref{eqn:nonBPScoset}).
Our considerations generalize this result to a moduli space of non-extremal KK black holes. The exact masslessness of moduli is
protected by the breaking of global symmetries so supersymmetry is not needed.

\subsection{The ${\cal N}=2$ Truncation}\label{sec:N=2 truncation}
Starting from ${\cal N}=4$ SUGRA with $n$ ${\cal N}=4$ matter multiplets, there is a consistent truncation to ${\cal N}=2$ SUGRA with
$n+1$ ${\cal N}=2$ vector multiplets that respects the KK black hole background. It is defined by keeping only fields that are even under the
$SU(4)_R$ element ${\rm diag} (I_2,-I_2)$.

All fermions, both gravitini and gaugini are in the fundamental ${\bf 4}$ of $SU(4)_R$ so the consistent truncation retains
exactly $1/2$ of them. In particular, the SUSY is reduced from ${\cal N}=4$ to ${\cal N}=2$. The bosons are either invariant under $SU(4)_R$ or
they transform as an antisymmetric tensor ${\bf 6}$. The branching rule ${\bf 6}\to 2(1,1)\oplus (2,2)$ under
$SU(4)_R\to SU(2)^2$ determines that its truncation retains only the $2$ fields on the skew-diagonal of the antisymmetric $4\times 4$ tensor.

The truncated theory has $2(2n+4)$ fermionic degrees of freedom and the same number of bosonic ones. We can implement
the truncation directly on the ${\cal N}=4$ coset (\ref{eqn:scalarcoset}) and find that scalars of the truncated theory parametrize
\begin{equation}
\frac{SU(1,1)}{U(1)}\times \frac{SO(2,n)}{SO(2)\times SO(n)}~.
\label{eqn:stncoset}
\end{equation}
This theory is known as the $ST(n)$ model. In the special case $n=2$ the $ST(2)$ model is the well-known $STU$ model. This model has
enhanced symmetry ensuring that its $3$ complex scalar fields are equivalent and similarly that its $4$ field strengths are equivalent.
The $STU$ model often appears as a subsector of more general ${\cal N}=2$ SUGRA theories, such as those defined by a cubic prepotential.
These in turn arise as the low energy limit of string theory compactified on a Calabi-Yau manifold, so the $STU$ model may capture some
generic features of such theories.

The consistent truncation to the $ST(n)$ model in ${\cal N}=2$ SUGRA is compatible with the embedding of the KK black hole in
${\cal N}=8$ SUGRA. The embedding (\ref{N=8:embedding}) in ${\cal N}=8$ excites precisely the field strengths on the skew-diagonal, breaking
$SU(8)_R\to USp(8)$. As discussed in (\ref{sec:N=4sec}), they were retained by the truncation to ${\cal N}=4$ SUGRA. The further truncation
of the antisymmetric representation to ${\cal N}=2$ SUGRA projects ${\bf 6}\to 2({\bf 1},{\bf  1})$ and so it specifically retains field strengths on the skew
diagonal. Moreover, the gauge fields that are projected out are in the ${\bf 2}$ of an $SU(2)$ so they are not coupled to other fields at quadratic order.

It can be shown that the ${\cal N}=4$ embedding identifies the ``dilaton" of the KK black hole with the scalar (as opposed to the pseudoscalar)
in the coset $SU(1,1)/U(1)$. This part of the scalar coset is untouched by the truncation to ${\cal N}=2$ SUGRA. Therefore, the truncation
to ${\cal N}=2$ does not remove any of the fields that are turned on in the background, nor any of those that couple to them at quadratic order.
This shows that the consistent truncation to ${\cal N}=2$ SUGRA, like other truncations considered in this section, removes only entire blocks of
fluctuations: the fields that remain have the same couplings as they do in the ${\cal N}=8$ context.

The breaking pattern determines the moduli space of scalars for the black hole background as
\begin{equation}
SU(1,1)\times \frac{SO(1,n-1)}{SO(n-1)}~.
\label{eqn:stnmoduli}
\end{equation}
In particular this confirms that, among the $2n+2$ scalars of the $ST(n)$ model, exactly $n$ are moduli and so are minimally
coupled massless scalars.

\subsection{More Comments on Consistent Truncations}
The natural endpoint of the consistent truncations is ${\cal N}=0$ SUGRA, {\it i.e.} the pure Kaluza-Klein theory (\ref{eqn:KKtheory}).
We constructed our embedding ({\ref{N=8:embedding}) into ${\cal N}=8$ SUGRA so that the Kaluza-Klein black hole would remain a
solution also to the full ${\cal N}=8$ SUGRA. Thus we arranged that all the additional fields required by ${\cal N}=8$ supersymmetry would be ``unimportant",  in the sense that they can be taken to vanish on the Kaluza-Klein black hole. It is therefore consistent to remove them again,
and that is the content of the ``truncation to ${\cal N}=0$ SUGRA".

From this perspective, the truncations considered in this section are intermediate stages between ${\cal N}=8$ and ${\cal N}=0$ in that only some of
the ``unimportant" fields are included. For each value of ${\cal N}=6, 4, 2$, the requirement that the Kaluza-Klein black hole is a solution largely
determines the truncation. The resulting embedding of the $STU$ model into ${\cal N}=8$ SUGRA is very simple, and possibly simpler than others that
appear in the literature, in that symmetries between fields in the $STU$ model are manifest even without performing any electromagnetic
duality.

Having analyzed the spectrum of fluctuations around Kaluza-Klein black holes in the context of SUGRA with
${\cal N}=8, 6, 4, 2$ (and even ${\cal N}=0$), it is natural to inquire about the situation for SUGRA with odd ${\cal N}$. Our
embeddings in ${\cal N}=6, 4, 2$ rely on the skew-diagonal nature of the embedding in ${\cal N}=8$ so they do not have any
generalizations to odd ${\cal N}$. This fact is vacuous for ${\cal N}=7$ SUGRA which automatically implies ${\cal N}=8$.
Moreover, it is interesting that ${\cal N}=3,5$ SUGRA do not have any non-BPS branch at all: all extremal black holes in these theories must
be BPS (they preserve supersymmetry) \cite{Andrianopoli:2006ub}. This may indicate that our examples exhaust a large class of
non-BPS embeddings.

\section{The General KK Black Hole in $\cN=2$ SUGRA}\label{sec:N=2sugra}

In this section, we start afresh with an {\it arbitrary} {solution} to the $D=4$ Kaluza-Klein theory (\ref{eqn:KKtheory}), such as the general Kaluza-Klein black hole
(\ref{EoM:PhiKK}-\ref{EoM:gKK}). We embed this solution into $\mathcal{N}=2$ SUGRA with a {\it general cubic prepotential} and analyze
the quadratic fluctuations around the background in this setting. Along the way we make additional assumptions that further decouple the fluctuations, and ultimately
specialize to a constant background dilaton and $ST(n)$ prepotential. In this case the final results of the direct computations will be consistent with those found
in section \ref{sec:N=2 truncation}, by truncation from ${\cal N}=8$ SUGRA, and summarized in section \ref{N=8:Summary}.

The setup in this section complements our discussion of the Kaluza-Klein black hole in $\mathcal{N}=8$ SUGRA and its truncations to $\mathcal{N}<8$ SUGRA.
Here we do not assume vanishing background dilaton $\Phi^{\text{(KK)}}=0$ from the outset and we consider more general theories.

\subsection{$\mathcal{N}=2$ SUGRA with Cubic Prepotential}

We first introduce $\mathcal{N}=2$ SUGRA. We allow for matter in the form of $n_{V}$ $\mathcal{N}=2$ vector multiplets with couplings encoded in a cubic prepotential
\begin{align}\label{pp}
F=\frac{1}{\kappa^{2}}\frac{d_{ijk}X^{i}X^{j}X^{k}}{X^{0}}~,
\end{align}
where $d_{ijk}$ is totally symmetric. We also include $n_{H}$ $\mathcal{N}=2$ hypermultiplets. The theory is described by the $\mathcal{N}=2$ SUGRA Lagrangian
\begin{eqnarray}\label{L}
e^{-1}\mathcal{L}^{(\mathcal{N}=2)}&=&\kappa^{-2}\left(\frac{R}{2}-\bar{\psi}_{i\mu}\gamma^{\mu\nu\rho}D_{\nu}\psi^{i}_{\rho}\right)-g_{\alpha\bar{\beta}}\partial^{\mu}z^{\alpha}\partial_{\mu}z^{\bar{\beta}}-\frac{1}{2}h_{uv}\partial_{\mu}q^{u}\partial^{\mu}q^{v}\nonumber\\
&&+\left(-\frac{1}{4}i\mathcal{N}_{IJ}F^{+I}_{\mu\nu}F^{+\mu\nu J}+F^{-I}_{\mu\nu}\,\text{Im}\,\mathcal{N}_{IJ}Q^{\mu\nu-J}\right.\nonumber\\
&&\left.-\frac{1}{4}g_{\alpha\bar{\beta}}\bar{\chi}^{\alpha}_{i}\slashed{D}\chi^{i\bar{\beta}}-\bar{\zeta}_{A}\slashed{D}\zeta^{A}+\frac{1}{2}g_{\alpha\bar{\beta}}\bar{\psi}_{i\mu}\slashed{\partial}z^{\alpha}\gamma^{\mu}\chi^{i\bar{\beta}}+\text{h.c.}\right) ~,
\end{eqnarray}
where
\begin{eqnarray}
F^{\pm}_{\mu\nu}&=&\frac{1}{2}\left(F_{\mu\nu} \pm \widetilde{F}_{\mu\nu}\right), \text{ with }\widetilde{F}_{\mu\nu}=-\frac{i}{2}\epsilon_{\mu\nu\rho\sigma}F^{\rho\sigma}~,\\
Q^{\mu\nu-J}&\equiv&\overline{\nabla}_{\bar{\alpha}}\bar{X}^{J}\left(\frac{1}{8}g^{\beta\bar{\alpha}}C_{\beta\gamma\delta}\bar{\chi}_{i}^{\gamma}\gamma^{\mu\nu}\chi^{\delta}_{j}\epsilon^{ij}+\bar{\chi}^{\bar{\alpha}i}\gamma^{\mu}\psi^{\nu j}\epsilon^{ij}\right)~\\ \notag
&&+X^{J}\left(\bar{\psi}^{\mu}_{i}\psi^{\nu}_{j}\epsilon^{ij}+\frac{1}{2}\kappa^{2}\bar{\zeta}^{A}\gamma^{\mu\nu}\zeta^{B}C_{AB}\right)~.
\end{eqnarray}
We follow the notations and conventions from \cite{Freedman:2012zz}. In particular, the $\chi^{\alpha}_{i}=P_{L}\chi^{\alpha}_{i},\alpha=1, \dots, n_{V}$ denote the physical gaugini and
$\zeta^{A}=P_{L}\zeta^{A},A=1, \dots, 2\,n_{H}$ denote the hyperfermions. The K\"{a}hler covariant derivatives are
\begin{align}
&\nabla_{\alpha}{X}^{I}=\left(\partial_{\alpha}+\frac{1}{2}\kappa^{2}\partial_{\alpha}\mathcal{K}\right)X^{I}~,\\
&\overline{\nabla}_{\bar{\alpha}}{X}^{I}=\left(\partial_{\bar{\alpha}}-\frac{1}{2}\kappa^{2}\partial_{\bar{\alpha}}\mathcal{K}\right)X^{I}~,
\end{align}
where the K\"{a}hler potential $\mathcal{K}$
\begin{align}
e^{-\kappa^{2}\mathcal{K}}=-i(X^{I}\bar{F}_{I}-F_{I}\bar{X}^{I})~,
\end{align}
with $F_I=\partial_I F = {\partial F\over\partial X^I}$.

The projective coordinates $X^{I}$ (with $I=0,\ldots,n_V$) are related to physical coordinates as $z^{i}=X^{i}/X^{0}$
(with $i=1,\ldots,n_V$). We split the complex scalars $z^{i}$ into real and imaginary parts
\begin{align}
z^{i}=x^{i}-iy^{i}~.
\end{align}
With cubic prepotential (\ref{pp}) we have
\begin{align}
g_{i\bar{j}}
=\partial_I \partial_{\bar J} {\cal K} = \k^{-2}\left(-\frac{3d_{ij}}{2d}+\frac{9d_{i}d_{j}}{4d^{2}}\right)~,
\end{align}
where we define
\begin{align}
d_{ij}\equiv d_{ijk}y^{k}~, \qquad d_{i}\equiv d_{ijk}y^{j}y^{k}~, \qquad d\equiv d_{ijk}y^{i}y^{j}y^{k}~.
\label{eqn:ddconstraints}
 \end{align}
Finally, the scalar-vector coupling are encoded in
\begin{align}
\mathcal{N}_{IJ}=\mu_{IJ}+i\nu_{IJ}~,
\end{align}
with
\begin{align}
\mu_{IJ}=\kappa^{-2}\begin{pmatrix} 2d_{ijk}x^{i}x^{j}x^{k} & -3d_{ijk}x^{j}x^{k}\\-3d_{ijk}x^{j}x^{k} & 6d_{ijk}x^{k} \end{pmatrix}~,\end{align}
and
\begin{align}
\nu_{IJ}=\kappa^{-2}\begin{pmatrix} -d+6d_{\l m}x^{\l}x^{m}-\frac{9}{d}(d_{\l}x^{\l})^2 & \frac{9}{d}(d_{\l}x^{\l})d_{i}-6d_{i\l}x^{\l}\\ \frac{9}{d}(d_{\l}x^{\l})d_{i}-6d_{i\l}x^{\l}& 6d_{ij}-\frac{9}{d}(d_{i}d_{j})\end{pmatrix}~.
\label{eqn:nuij}
\end{align}

 \subsection{The Embedding into $\mathcal{N}=2$ SUGRA}
\label{sec:embedding}
We want to embed our seed solution into $\mathcal{N}=2$ SUGRA. The starting point is a solution to the equations of motion (\ref{EoM:PhiKK}, \ref{EoM:AKK}, \ref{EoM:gKK}) of the Kaluza-Klein theory. We denote the corresponding fields $g_{\mu\nu}^{(\text{KK})}$, $F^{(\text{KK})}_{\mu\nu}$
and $\Phi^{(\text{KK})}$. The fields of $\mathcal{N}=2$ SUGRA are then defined to be
\begin{align}\label{embedding}
&g_{\mu\nu}^{(\text{SUGRA})}=g_{\mu\nu}^{(\text{KK})}\nonumber ~,\\
&F^{0}_{\mu\nu}=\frac{1}{\sqrt{2}}F^{(\text{KK})}_{\mu\nu},\quad F^{i}_{\mu\nu}=0,\quad \text{for } 1\leq i\leq n_{V}\nonumber\\
&x^{i}=0,\quad \text{for } 1\leq i\leq n_{V}\nonumber~,\\
&y^{i}=c^{i}y_{0}, \text{ with } y_{0}=\frac{\exp\left(-2\Phi^{(\text{KK})}/\sqrt{3}\right)}{(d_{ijk}c^{i}c^{j}c^{k})^{1/3}}\nonumber~,\\
&(\text{All other bosonic fields in }\mathcal{N}=2\text{ SUGRA})=0\nonumber~,\\
&(\text{All fermionic fields in }\mathcal{N}=2\text{ SUGRA})=0~.
\end{align}
This field configuration solves the equations of motion of $\mathcal{N}=2$ SUGRA for any seed solution
to the Kaluza-Klein theory. In the following, we will often declutter formulae by omitting the superscript ``KK" when referring to fields in the seed solution.

The embedding (\ref{embedding}) is really a family of embeddings parameterized by the $n_V$ constants
$c^{i}$ (with $i=1,\ldots, n_V$). They are projective coordinates on the moduli space parametrized by the $n_V$ scalar fields
$y_i$ with the constraint
\begin{equation}
d = d_{ijk} y^i y^j y^k = \exp\left(-2\sqrt{3}\Phi^{(\text{KK})}\right)~.
\label{eqn:ddef}
\end{equation}
In the special case of the non-rotating Kaluza-Klein black hole with $P=Q$, we have $\Phi^{(\text{KK})}=0$ and so the constraint is $d=1$.
More generally, $d$ is the composite field defined through the constraints (\ref{eqn:ddconstraints}) and related to the Kaluza-Klein dilaton by (\ref{eqn:ddef}).

\subsection{Decoupled Fluctuations: General Case}
The Lagrangian for quadratic fluctuations around a bosonic background always decouples into a bosonic sector and fermionic sector,
\begin{eqnarray}
\delta^{2}\mathcal{L}^{(\mathcal{N}=2)}&=&\delta^{2}\mathcal{L}^{(\mathcal{N}=2)}_{\text{bosons}}+\delta^{2}\mathcal{L}^{(\mathcal{N}=2)}_{\text{fermions}}~.
\end{eqnarray}
With the above embedding into ${\mathcal N=2}$, each sector further decouples into several blocks.

\medskip

The bosonic sector decomposes as the sum of three blocks
\begin{eqnarray}
\delta^{2}\mathcal{L}^{(\mathcal{N}=2)}_{\text{bosons}}&=&\delta^{2}\mathcal{L}^{(\mathcal{N}=2)}_{\text{gravity}}+\delta^{2}\mathcal{L}^{(\mathcal{N}=2)}_{\text{vectors}}+\delta^{2}\mathcal{L}^{(\mathcal{N}=2)}_{\text{scalars}}~.
\end{eqnarray}
The ``gravity block" $\delta^{2}\mathcal{L}^{(\mathcal{N}=2)}_{\text{gravity}}$ consists of the graviton $\delta g_{\mu\nu}$, the gauge field $\delta A^{0}_{\mu}$, and
the $n_{V}$ real scalars $\delta y^{i}$:
\begin{eqnarray}\label{N=2:b1}
e^{-1}\delta^{2}\mathcal{L}^{(\mathcal{N}=2)}_{\text{gravity}}&=&\frac{1}{\sqrt{-g}}\delta^{2}\left[\sqrt{-g}\left(\frac{R}{2\kappa^{2}}-g_{ij}\partial_{\mu} y^{i}\partial^{\mu} y^{j}
+\frac{d}{4\kappa^{2}}F^{0}_{\mu\nu}F^{\mu\nu 0}\right)\right]~.
\end{eqnarray}
Generically, the fields $\delta g_{\mu\nu},\delta A^{0}_{\mu}$ and $\delta y^{i}$ all mix together. This block can nonetheless be further decoupled with simplifying assumptions, as we will discuss later.

The block $\delta^{2}\mathcal{L}^{(\mathcal{N}=2)}_{\text{vectors}}$ consists of the $n_{V}$ vector fields $\delta A^{i}_{\mu}$ and the $n_{V}$ real
pseudoscalars $\delta x^{i}$:
\begin{eqnarray}
e^{-1}\delta^{2}\mathcal{L}^{(\mathcal{N}=2)}_{\text{vectors}}&=&g_{ij}\left(-\partial_{\mu}\delta x^{i}\partial^{\mu}\delta x^{j}-\frac{1}{2}dF_{\mu\nu}F^{\mu\nu}\delta x^{i}\delta x^{j}+\sqrt{2}dF_{\mu\nu}\delta x^{i}\delta F^{\mu\nu j}-d\delta F^{i}_{\mu\nu}\delta F^{\mu\nu j}\right)\nonumber~.
\end{eqnarray}
The K\"{a}hler metric $g_{ij}$ can be diagonalized and we obtain $n_{V}$ identical decoupled copies, that we call ``vector block", each consisting in one vector field and one real scalar. Denoting the fluctuating field $f_{\mu\nu}$, one such copy has the Lagrangian
\begin{align}\label{N=2:b2}
e^{-1}\delta^{2}\mathcal{L}^{(\mathcal{N}=2)}_{\text{vector}} &
=- {1\over 2} \partial_{\mu} x\partial^{\mu} x-\frac{d}{4} F_{\mu\nu}F^{\mu\nu} x^2+\frac{d}{2} F_{\mu\nu} f^{\mu\nu} x
-{d\over 4} f_{\mu\nu}f^{\mu\nu}~,
\end{align}
using conventional normalizations for the scalar fields.
%

The last bosonic block contains the hyperbosons:
\begin{eqnarray}\label{N=2:b3}
e^{-1}\delta^{2}\mathcal{L}^{(\mathcal{N}=2)}_{\text{scalars}}&=&-\frac{1}{2}h_{uv}\partial_{\mu}\delta q^{u}\partial^{\mu}\delta q^{v}~.
\end{eqnarray}
The quaternionic K\"{a}hler metric $h_{uv}$ is trivial on the background. Hence, this block decouples at quadratic order into $4\,n_{H}$ independent minimally coupled massless scalars.

\medskip

We next turn to the fermions. The Lagrangian (\ref{L}) is the sum of the decoupled Lagrangians
\begin{eqnarray}
\delta^{2}\mathcal{L}^{(\mathcal{N}=2)}_{\text{fermions}}=\delta^{2}\mathcal{L}^{(\mathcal{N}=2)}_{\text{hyperfermions}}+\delta^{2}\mathcal{L}^{(\mathcal{N}=2)}_{\text{gravitino-gaugino}}~.
\end{eqnarray}
The hyperfermions consist of $n_{H}$ identical copies, that we call ``hyperfermion block", each containing two hyperfermions. For any two such fermions we
can take $C_{AB}=\epsilon_{AB}$ with $A,B=1,2$. The resulting Lagrangian is
\begin{eqnarray}
    e^{-1}\delta^{2}\mathcal{L}^{(\mathcal{N}=2)}_{\text{hyperfermion}}&=&-2\bar{\zeta}_{A}\slashed{D}\zeta^{A}+\left(\frac{\kappa^{2}}{2}F^{-I}_{\mu\nu}\nu_{IJ}X^{J}\bar{\zeta}^{A}\gamma^{\mu\nu}\zeta^{B}\epsilon_{AB}+\text{h.c.}\right)~.
\end{eqnarray}
In our background, we use  (\ref{embedding}, \ref{eqn:nuij}) to find
\begin{eqnarray}\label{N=2:hyperfermion}
e^{-1}\delta^{2}\mathcal{L}^{(\mathcal{N}=2)}_{\text{hyperfermion}}&=&-2\bar{\zeta}_{A}\slashed{D}\zeta^{A}-\left(\frac{d^{\frac{1}{2}}}{8} F^{-}_{\mu\nu}\bar{\zeta}^{A}\gamma^{\mu\nu}\zeta^{B}\epsilon_{AB}+\text{h.c.}\right)~.
\end{eqnarray}
We used  the $T$-gauge \cite{Freedman:2012zz} to  fix the projective coordinates $X^{I}$ resulting in $X^{0}=(8d)^{-1/2}$.

The ``gravitino-gaugino block" contains two gravitini and $n_V$ gaugini and has Lagrangian
\begin{eqnarray}\label{N=2:gravitino+gaugino}
e^{-1}\delta^{2}\mathcal{L}^{(\mathcal{N}=2)}_{\text{gravitino-gaugino}}&=&-\frac{1}{\kappa^{2}}\bar{\psi}_{i\mu}\gamma^{\mu\nu\rho}D_{\nu}\psi^{i}_{\rho}+\left(-\frac{d^{\frac{1}{2}}}{4\kappa^{2}}F^{-}_{\mu\nu}\bar{\psi}^{\mu}_{i}\psi^{\nu}_{j}\epsilon^{ij}\right.\nonumber\\
&&+\frac{9}{256\kappa^{2}d^{\frac{3}{2}}}F^{-}_{\mu\nu}d_{\bar{\alpha}}g^{\beta\bar{\alpha}}d_{\beta\gamma\delta}\bar{\chi}_{i}^{\gamma}\gamma^{\mu\nu}\chi^{\delta}_{j}\epsilon^{ij}-\frac{3i}{8\kappa^{2}d^{\frac{1}{2}}}F^{-}_{\mu\nu}d_{\bar{\alpha}}\bar{\chi}^{\bar{\alpha}i}\gamma^{\mu}\psi^{\nu j}\epsilon^{ij}\nonumber\\
&&\left.-\frac{1}{4}g_{\alpha\bar{\beta}}\bar{\chi}^{\alpha}_{i}\slashed{D}\chi^{i\bar{\beta}}+\frac{1}{2}g_{\alpha\bar{\beta}}\bar{\psi}_{ia}\slashed{\partial}z^{\alpha}\gamma^{a}\chi^{i\bar{\beta}}+\text{h.c.}\right)~.
\end{eqnarray}
Generally, all the gravitini and gaugini couple nontrivially but they can be further decoupled in simpler cases, as we will discuss later.
\medskip

Summarizing so far: given any Kaluza-Klein solution, the embedding (\ref{embedding}) provides solutions of $\mathcal{N}=2$ SUGRA. We have expanded the $\mathcal{N}=2$ Lagrangian around this background to quadratic order and observed that the fluctuations can be decoupled as shown in Table \ref{N=2:FluctTable}.
%
%
%
\begin{table}[H]
	\centering
\begin{tabular}{|c|c|c|c|c|}
	\hline
Degeneracy &	Multiplet 	& Block content &  Lagrangian \\\hline\hline
	1 & Gravity block & 1 graviton, 1 vector, $n_{V}$ scalars & (\ref{N=2:b1}) \\\hline
		$n_{V}$ &	Vector block  & 1 vector and 1 (pseudo)scalar & (\ref{N=2:b2})  \\\hline
$4n_{H}$ & Scalar block & 1 real scalar & (\ref{N=2:b3})  \\\hline
1 &	Gravitino-gaugino block &2 gravitini and $2\,n_{V}$ gaugini  & (\ref{N=2:gravitino+gaugino})  \\\hline
$n_{H}$ & 	Hyperfermion block &2 hyperfermions & (\ref{N=2:hyperfermion}) \\
	\hline
\end{tabular}
	
	\caption{ Decoupled quadratic fluctuations in $\cN=2$ SUGRA around a general KK black hole.}\label{N=2:FluctTable}
\end{table}
These results are reminiscent of the analogous structure for ${\cal N}=8$ SUGRA, summarized in (\ref{N=8:embedding}). However, with the more general assumptions made here, there are more scalars in the $\cN=2$ gravity block than in the analogous $\cN=8$ KK block and these additional scalars do not generally
decouple from gravity. Similarly, the $\cN=2$ gravitino-gaugino block here includes more gaugini than the analogous $\cN=8$ gravitino block.


\subsection{Decoupled Fluctuations: Constant Dilaton}\label{N=2:bosons}

So far, we have been completely general about the underlying Kaluza-Klein solution. In this section, we further decouple the quadratic fluctuations by assuming that the scalar fields of $\mathcal{N}=2$ SUGRA
are constant
\begin{align}
\qquad y^{i}=\text{constant},\ i=1,...,n_{V}~.
\end{align}
From the embedding (\ref{embedding}), this is equivalent to taking the Kaluza-Klein dilaton to vanish
\be
\Phi^{(\text{KK})}=0~,
\ee
since we can always rescale the field strengths to arrange for $d=d_{ijk}y^{i}y^{j}y^{k}=1$. As noted previously, this is satisfied by the non-rotating Kaluza-Klein black hole with $P=Q$. This is the simplified background that we already studied in ${\cal N}=8$ SUGRA, but it is embedded here in ${\cal N}=2$ SUGRA with arbitrary prepotential. As in the $\cN=8$ case, we will use that the background satisfies
\begin{align}
R=0~, \qquad F_{\mu\nu}F^{\mu\nu}=0~
\end{align}
to decouple further the quadratic fluctuations.

\begin{itemize}
	
	\item \emph{Gravity}	
	
The gravity block decouples as
\begin{align}
\delta^{2}\mathcal{L}^{(\mathcal{N}=2)}_{\text{gravity}}=\delta^{2}\mathcal{L}^{(\mathcal{N}=2)}_{\text{KK}}+\delta^{2}\mathcal{L}^{(\mathcal{N}=2)}_{\text{relative}}~,
\end{align}
where $\delta^{2}\mathcal{L}^{(\mathcal{N}=2)}_{\text{KK}}$ is the ``KK block", consisting of the graviton $\delta g_{\mu\nu}$, the graviphoton $\delta A^{0}_{\mu}$ and the center-of-mass scalar
$\delta y'^{1}$. $\delta^{2}\mathcal{L}_{\text{relative}}^{(\mathcal{N}=2)}$ denotes $n_{V}-1$ free massless scalars $\delta y'^{i},\ i=2,\ldots n_V$. This decoupling is obtained by center-of-mass diagonalization: the $\delta y'^{i}$ are linear combinations of $\delta y^{i}$ such that $\delta y'^{1}$ is precisely the combination that couples to the graviton and graviphoton at quadratic order.  Then, the ``relative scalars" $\delta y'^{i},\ i=2,\dots,n_V$ are minimally coupled to the background
\begin{eqnarray}
\label{yrel}
e^{-1}\delta^{2}\mathcal{L}^{(\mathcal{N}=2)}_{\text{relative}}&=&-\frac{2}{\kappa^{2}}\partial_{\mu}\delta y'^{i}\partial^{\mu}\delta y'^{i}\quad (\text{for }i=2,\dots,n_{V})
~,
\end{eqnarray}
The center-of-mass Lagrangian turns out to be exactly the same as the $\cN=8$ KK block (\ref{N=8:CMb})
\begin{align}\label{N=2:CMb}
\delta^{2}\mathcal{L}^{(\mathcal{N}=2)}_{\text{KK}}=&~\delta^{2}\mathcal{L}^{(\mathcal{N}=8)}_{\text{KK}} ~,
\end{align}
with the identifications
\begin{align}
&\bar{h}_{\mu\nu}=\frac{1}{\sqrt{2}}\left(\delta g_{\mu\nu}-\frac{1}{4}g_{\mu\nu}g^{\rho\sigma}\delta g_{\rho\sigma}\right)~, \quad h=\frac{1}{\sqrt{2}}g^{\rho\sigma}\delta g_{\rho\sigma}~,\\
&a_{\mu}=\sqrt{2}\delta A^{0}_{\mu}~, \quad  f_{\mu\nu}=\partial_{\mu}a_{\nu}-\partial_{\nu}a_{\mu}~,\\
&\phi=\delta y'^{1}=-\frac{\sqrt{3}d_{i}}{2d}\delta y^{i}=\delta\Phi~.
\end{align}
The equality between $\delta^{2}\mathcal{L}^{(\mathcal{N}=2)}_{\text{KK}}$ and $\delta^{2}\mathcal{L}^{(\mathcal{N}=8)}_{\text{KK}}$ is expected because the KK block is the same for any $\cN=2$ SUGRA and in particular for the $\cN=2$ truncations of $\cN=8$ SUGRA.

The $n_{V}-1$ minimally coupled massless scalars $\delta y'^{i}, i=2, \dots, n_{V}$ parameterize flat directions in the moduli space, at least at quadratic order. In important situations with higher symmetry, including homogeneous spaces constructed as coset manifolds, it can be shown that these $n_{V}-1$ directions are exactly flat at all orders. This implies that, in particular, these models are stable \cite{Tripathy:2005qp,Bellucci:2006xz}. In such situations the ``relative" coordinates $\delta y'^{i}$ are Goldstone bosons parameterizing symmetries of the theories.

\item \emph{Vector block}

Using the fact that $F_{\mu\nu}F^{\mu\nu}=0$, the vector block becomes
\begin{align}\label{N=2:vector}
e^{-1}\delta^{2}\mathcal{L}^{(\mathcal{N}=2)}_{\text{vector}} &
=-\frac{1}{2} \partial_{\mu} x\partial^{\mu} x+\frac{1}{2} F_{\mu\nu} f^{\mu\nu} x
-\frac{1}{4} f_{\mu\nu}f^{\mu\nu}~.
\end{align}
Again, we find that  $\delta^{2}\mathcal{L}^{(\mathcal{N}=2)}_{\text{vector}}=\delta^{2}\mathcal{L}^{(\mathcal{N}=8)}_{\text{vector}}$ after proper normalization of the field strength.

\item \emph{Scalar block}

The Lagrangian for hyperbosons $\delta^{2}\mathcal{L}^{(\mathcal{N}=2)}_{\text{scalars}}$ consists of $4 n_H$ minimally coupled scalars. In addition, the center-of-mass diagonalization has brought $n_V-1$ minimally coupled ``relative" scalars $\delta^{2}\mathcal{L}^{(\mathcal{N}=2)}_{\text{relative}}$. This gives a total of $n_{V}+4 n_{H}-1$ minimally coupled scalars.

\end{itemize}

We now turn to fermions. The interactions between gravitini and gaugini simplify greatly when scalars are constant. However, they still depend on the prepotential through the structure constants
$d_{\alpha\beta\gamma}$. The fermionic fluctuations in ${\cal N}=2$ SUGRA are therefore qualitatively different from the bosonic fluctuations which,
as we just saw, reduce to the form found in ${\cal N}=8$ SUGRA.

For fermions we need to further specialize and study the $ST(n)$ model. This model already appeared in section \ref{sec:N=2 truncation},
as a truncation of ${\cal N}=8$ SUGRA to ${\cal N}=2$. Presently, we introduce it as the model with $n_{V}=n+1$ vector multiplets and prepotential
\begin{align}
F=\frac{1}{\kappa^{2}}\frac{X^{1}(X^{2}X^{2}-X^{\alpha}X^{\alpha})}{2X^{0}}\qquad (\alpha=3, \dots,n_{V})~.
\end{align}
We take the background scalars
\begin{align}\label{ybstn}
y^{1}=1,\quad y^{2}=\sqrt{2},\quad y^{\alpha}=0\quad (\alpha=3, \dots,n_{V})~,
\end{align}
such that the normalization is $d=1$ and therefore $\Phi^{\text{(KK)}}=0$. As mentioned already in section \ref{sec:N=2 truncation}, this model
generalizes the $STU$ model which is equivalent to $ST(2)$.

\begin{itemize}
\item \emph{Gravitino-gaugino block}

The Lagrangian for the gravitino-gaugino block decouples as
\begin{eqnarray}
\delta^{2}\mathcal{L}^{(\mathcal{N}=2)}_{\text{gravitino-gaugino}}&=&\delta^{2}\mathcal{L}^{(\mathcal{N}=2)}_{\text{gravitino}}+\delta^{2}\mathcal{L}^{(\mathcal{N}=2)}_{\text{gaugino}}~,
\end{eqnarray}
after using center-of-mass diagonalization. We call $\chi^{\prime i 1}$ the center-of-mass gaugini, {\it i.e.} the gaugini that couples to the gravitini. More precisely, we define
\begin{align}
&\chi'^{i1}=\frac{1}{4}\le(\frac{\sqrt{3}}{3}\chi^{i1}+\frac{\sqrt{6}}{3}\chi^{i2}\ri)~, \quad \chi'^{i2}=\frac{1}{4}\le(\frac{\sqrt{6}}{3}\chi^{i1}-\frac{\sqrt{3}}{3}\chi^{i2}\ri)~,\nonumber\\
&\chi'^{i\alpha}=\frac{1}{4}\chi^{i\alpha} \quad \text{ for }\alpha=3, \dots, n_{V}~.
\end{align}
We find a center-of-mass multiplet that we call ``gravitino block"
\begin{eqnarray}\label{N=2:CMf}
e^{-1}\delta^{2}\mathcal{L}^{(\mathcal{N}=2)}_{\text{gravitino}}&=&-\frac{1}{\kappa^{2}}\bar{\psi}_{i\mu}\gamma^{\mu\nu\rho}D_{\nu}\psi^{i}_{\rho}+\frac{1}{\kappa^{2}}\left(-\bar{\chi}'^{1}_{i}\slashed{D}\chi'^{i1}
-\frac{1}{4}\bar{\psi}^{\mu}_{i}F^{-}_{\mu\nu}\psi^{\nu}_{j}\epsilon^{ij}\right.\nonumber\\
&&\left.+\frac{1}{4}\bar{\chi}'^{1}_{i}F^{-}_{\mu\nu}\gamma^{\mu\nu}\chi'^{1}_{j}\epsilon^{ij}-\frac{\sqrt{3}i}{2}\bar{\chi}'^{i1}\gamma^{\mu}F^{-}_{\mu\nu}\psi^{\nu j}\epsilon^{ij}+\text{h.c.}\right)~,
\end{eqnarray}
This Lagrangian couples the two gravitini to two center-of-mass gaugini. The ``relative" multiplets are $n_{V}-1$ identical copies of a ``gaugino block"
\begin{eqnarray}
\label{N=2:gaugino}
e^{-1}\delta^{2}\mathcal{L}^{(\mathcal{N}=2)}_{\text{gaugino}}&=&-\frac{2}{\kappa^{2}}\bar{\chi}'^{\alpha}_{i}\slashed{D}\chi'^{i}_{\alpha}-\left(\frac{1}{8\kappa^{2}}\bar{\chi}'^{\alpha}_{i}F^{-}_{\mu\nu}\gamma^{\mu\nu}\chi'_{j\alpha}\epsilon^{ij}+\text{h.c.}\right)~,
\end{eqnarray}
where $\alpha=2,\dots,n_{V}$.

\item \emph{Hyperfermion block}

The hyperfermion Lagrangian is given in (\ref{N=2:hyperfermion}). We notice that
\begin{equation}
\delta^{2}\mathcal{L}^{(\mathcal{N}=2)}_{\text{hyperfermion}} = \delta^{2}\mathcal{L}^{(\mathcal{N}=2)}_{\text{gaugino}}~,
\label{eqn:gauginihyper}
\end{equation}
The fluctuations of ``relative" gaugini are therefore the same as the fluctuations of hyperfermions. Therefore, we call both of them ``gaugino block".

\end{itemize}

The Lagrangians (\ref{N=2:CMf}) and (\ref{N=2:gaugino}) are written in terms of Weyl fermions. If we rewrite them with Majorana fermions, we find that
\begin{align}
\delta^{2}\mathcal{L}^{(\mathcal{N}=2)}_{\text{gravitino}} &=\delta^{2}\mathcal{L}^{(\mathcal{N}=8)}_{\text{gravitino}} ~,\\
\delta^{2}\mathcal{L}^{(\mathcal{N}=2)}_{\text{gaugino}}&=\delta^{2}\mathcal{L}^{(\mathcal{N}=8)}_{\text{gaugino}}~,
\end{align}
where the right-hand sides were defined in (\ref{N=8:gravitino}) and  (\ref{N=8:gaugino}). The agreement between our explicit computations of the fermionic
blocks for the $ST(n)$ model in $\mathcal{N}=2$ SUGRA and the analogous results in $\mathcal{N}=8$ SUGRA is an important consistency check on
the truncations discussed in section \ref{sec:N=2 truncation}. This also explains the agreement (\ref{eqn:gauginihyper}) between
fermionic fluctuations that are  in different ${\cal N}=2$ multiplets.
${\cal N}=2$ gaugini and hyperfermions becomes equivalent when embedded into some larger structure, ultimately furnished by ${\cal N}=8$ SUGRA.

\medskip

In summary, taking the dilaton to be constant has further decoupled the fluctuations in $\cN=2$ SUGRA around the KK background, as shown in Table \ref{N=2:ConstFluctTable}. For bosons, we recover the results of $\cN=8$ SUGRA as expected, although we are more general here since we allow for an arbitrary prepotential. For fermions, we have to specialize to the $ST(n)$ model to be able to further decouple the fluctuations. The resulting fermionic fluctuations also reproduce the fluctuations of $\cN=8$ SUGRA.

\begin{table}[H]
	\centering
	\begin{tabular}{|c|c|c|c|c|}
		\hline
		Degeneracy &	Multiplet 	& Block content &  Lagrangian \\\hline\hline
		1 & KK block & 1 graviton, 1 vector, 1 scalar & (\ref{N=2:CMb}) \\\hline
		$n_{V}$ &	Vector block  & 1 vector and 1 (pseudo)scalar & (\ref{N=2:vector})  \\\hline
		$n_{V}+4n_{H}-1$ & Scalar block & 1 real scalar & (\ref{N=2:b3}, \ref{yrel}) \\\hline
		1 &	Gravitino block &2 gravitini and 2 gaugini  & (\ref{N=2:CMf})  \\\hline
		$n_{V}+n_{H}-1$ & 	Gaugino block &2 spin $1/2$ fermions & (\ref{N=2:hyperfermion}, \ref{N=2:gaugino}) \\
		\hline
	\end{tabular}
	
	\caption{ Decoupled fluctuations in $\cN=2$ SUGRA around the KK black hole with constant dilaton. The decoupling in the bosonic sector holds for an arbitrary prepotential. The fermionic sector has been further  decoupled by specializing to the $ST(n)$ model. }\label{N=2:ConstFluctTable}
\end{table}

\section{Logarithmic Corrections to Black Hole Entropy}
\label{sec:lognonbps}

The logarithmic correction controlled by the size of the horizon in Planck units is computed by the functional determinant of the quadratic fluctuations of
light fields around the background solution. The arguments establishing this claim for non-extremal black holes are made carefully in \cite{Sen:2012dw}.
In this section we give a brief summary of the steps needed to extract the logarithm using the heat kernel approach. It follows the discussion
in \cite{Charles:2015eha} and we refer to \cite{Vassilevich2003} for background literature on technical aspects.

Naturally, we apply the procedure to the Kaluza-Klein black holes on the non-BPS branch. This gives our final results for the coefficients of the logarithmic corrections, summarized in Table \ref{tab:degeneracy1}.

\subsection{General Framework: Heat Kernel Expansion}

In Euclidean signature, the effective action $W$ for the quadratic fluctuations takes the schematic form
\be\label{eqn:effaction}
e^{-W} = \int {\cal D} \phi \exp\left({-}\int d^4x \sqrt{g}\,\phi_n \Lambda_m^n \phi^m\right){=\det}^{\mp 1/2}\Lambda~,
\ee
where $\Lambda$ is a second order differential operator that characterizes the background solution, and $\phi_n$ embodies the entire field content of the theory.
The sign $\mp$ is $-$ for bosons and $+$ for fermions.  The formal determinant of $\Lambda$ diverges and a canonical way to regulate it is by introducing a heat kernel: if $\{\lambda_i\}$ is the set of eigenvalues of $\Lambda$, then the heat kernel $D(s)$ is defined by
\be\label{eq:dsl}
D(s)=\Tr\:e^{-s\Lambda}=\sum_{i}e^{-s\lambda_i}~,
\ee
and the effective action becomes
\be\label{eq:w1}
W=\mp {1\/2}\int_{\epsilon}^\infty {ds\over s} D(s)~.
\ee
Here $\epsilon$ is an ultraviolet cutoff, which is typically controlled by the Planck length, {\it i.e.} $\epsilon\sim \ell_{P}^2\sim G$.

In our setting it is sufficient to focus on the contribution of massless fields in the two derivative theory. For this part of the spectrum, the
scale of the eigenvalues $\lambda_i$ is set by the background size which in our case is identified with the size of the black hole horizon, denoted by $A_H$.
The integral (\ref{eq:w1}) is therefore dominated by the integration range $\epsilon\ll s\ll A_H$, and there is a  logarithmic contribution
\be\label{eq:expansion}
\int_{\epsilon}^\infty {ds\over s} D(s) = \cdots + C_{\rm local}  \log(A_H/G)+\cdots~.
\ee
with coefficient denoted by $C_{\rm local}$.
This term comes from the constant term in the Laurent expansion of the heat kernel $D(s)$. Introducing the heat kernel density $K(x,x;s)$ which satisfies
\be
D(s)= \int d^4x \sqrt{g}\, K(x,x;s)~,
\ee
it is customary to cast the perturbative expansion in $s$ as
\be
K(x,x;s)=\sum_{n=0}^\infty s^{n-2} a_{2n}(x)~,
\ee
and we identify
\begin{align}
	C_{\text{local}} = \int d^4 x \sqrt{g} \, a_4(x)~.
\end{align}
The functions $\{a_{2n}(x)\}$ are known as the Seeley-DeWitt coefficients. The logarithmic term that we need is controlled by $a_4(x)$.
The omitted terms denoted by ellipses in \eqref{eq:expansion} are captured by the other Seeley-DeWitt coefficients. For example, the term $a_0(x)$ induces
a cosmological constant at one-loop and the term $a_2(x)$ renormalizes Newton constant.

There is a systematic way to evaluate the Seeley-DeWitt coefficients in terms of the background fields and covariant derivatives appearing in the operator $\Lambda$ \cite{Vassilevich2003}. The procedure assumes that the quadratic fluctuations can be cast in the form
\be\label{eq:diff}
{-}\L^n_m = (\Box) I^n_m + 2 (\w^\mu D_\mu)^n_m + P^n_m~.
\ee
Here, $I^n_m$ is the identity matrix in the space of fields, $\omega^\mu$ and $P$ are matrices constructed from the background fields, and $\Box=D_\mu D^\mu$. From this data, the Seeley-DeWitt coefficient $a_4(x)$ is given by the expression
\be\label{a4}
(4\pi)^2 a_4(x) = \Tr\left[{1\over2} E^2 + {1\over6} R E + {1\over 12} \Om_\mn\Om^\mn + {1\over360} (5 R^2 + 2R_{\mn\rs}R^{\mn\rs} - 2R_\mn R^\mn)\right]~,
\ee
where
\be
E = P - \w^\mu\w_\mu - (D^\mu\w_\mu) ~, \qquad \Om_\mn = [D_\mu+\w_\mu,D_\nu+\w_\nu]~.
\ee
This is the advantage of the heat kernel approach: after explicitly expanding the action around the background to second order,
we have a straightforward formula to compute the Seeley-DeWitt coefficients from $\Lambda$ (\ref{eq:diff}).

The preceding discussion is based on the operator $\Lambda$ (\ref{eq:diff}) that is second order in derivatives. For fermions, the quadratic fluctuations are
described by a first order operator $H$ so the discussion must be modified slightly. We express the quadratic Lagrangian as
\be
\delta^2 \cL = \bar{\Psi} H \Psi~.
\ee
Following the conventions in \cite{Charles:2015eha}, we always cast the quadratic fluctuations for the fermions in terms of Majorana spinors.
The one-loop action is obtained by applying heat kernel techniques to the operator $H^\dg H$ and using
\be
\log \det H = {1\/2}\log \det H^\dg H~.
\ee
Fermi-Dirac statistics also gives an additional minus sign. Thus, the fermionic contribution is obtained by multiplying $(\ref{a4})$ with an additional factor of $-1/2$.

\subsection{Local Contributions}

It is conceptually straightforward to compute $a_4(x)$ via (\ref{a4}). However, it can be cumbersome to decompose the differential operators, write them in the form \eqref{eq:diff} and compute their traces. The main complication is that our matter content is not always minimally coupled, as
emphasized in sections \ref{sec:fluctuations} and \ref{sec:N=2sugra}.

To overcome these technical challenges we automated the computations using Mathematica with the symbolic tensor manipulation
package \ttt{xAct}\footnote{\href{http://www.xact.es/}{www.xact.es}}. In particular, we
used the subpackage \ttt{xPert}  \cite{Brizuela:2008ra} to expand the bosonic Lagrangian to second order. We created our own package  for treatment of Euclidean spinors. The computation proceeds as follows:
\begin{enumerate}
	\item Expand the Lagrangian to second order.
	\item Gauge-fix and identify the appropriate ghosts.
	\item Reorganize the fluctuation operator $\L_m^n$ and extract the operators $\omega_\mu$ and $P$ from \eqref{eq:diff}. 	
	\item Compute the Seeley-DeWitt coefficient $a_4(x)$ using formula (\ref{a4}).
	\item Simplify $a_4(x)$ using the background equations of motion, tensor and gamma matrix identities.
\end{enumerate}
The results of the expansion to second order with $\ttt{xPert}$ match with the bosonic Lagrangians summarized in Table \ref{FluctTable}. In Appendix \ref{app:a4} we elaborate on the intermediate steps and record the traces of $E$ and $\Om_{\mn}$ for each of the
blocks encountered in our discussion.

{\it A priori}, the Seeley-DeWitt coefficient $a_4(x)$ is a functional of both the geometry and the matter fields. The fact that the dilaton  $\Phi^{(\text{KK})}$ is constant on our background simplifies the situation greatly. By using the equations of motion, $a_4(x)$ can be recast as a functional of the geometry alone. We list the equations that we use to simplify $a_4(x)$ explicitly in Appendix \ref{app:a4}.

As a result, for our background, the Seeley-DeWitt coefficient at four derivative order can be arranged in the canonical form
\be\label{eq:a4min}
 a_4(x) = {c\over 16\pi^2} W_{\mu\nu\rho\sigma}W^{\mu\nu\rho\sigma}-{a\over 16\pi^2} E_4~,
\ee
where $a$ and $c$ are constants governed by the couplings and field content of the theory and the curvature invariants are
defined in \eqref{eq:w2} and \eqref{eq:e4}. The values of $c$ and $a$ are summarized in Tables \ref{KKres} and \ref{tab:degeneracy1}.

\begin{table}[H]
	\centering
\begin{tabular}{|c||c|c|c|c|c|}
\hline
Multiplet $\backslash$ Properties	& Content &d.o.f.& $c$ &  $a$ & $c-a$ \\\hline\hline
 Minimal boson & 1 real scalar& 1& ${1\/120}$ & ${1\/360}$ & ${1\/180}$ \\\hline
 Gaugino block &2 gaugini& 4&${13\/960}$ & $-{17\/2880}$ &  ${7\/360}$ \\\hline
Vector block  & 1 vector and 1 (pseudo)scalar& 3& ${1\/40}$ & ${11\/120}$ & $-{1\/ 15}$  \\\hline
Gravitino block &2 gravitini and 2 gaugini & 8&  $-{347\/480}$ & $-{137\/1440}$ & $-{113\/180}$ \\\hline
KK block & 1 graviton, 1 vector, 1 scalar& 5& ${37\/24}$ & ${31\/72}$ &  ${10\/9}$ \\
\hline
	\end{tabular}
	
	\caption{ Contributions to $a_4(x)$ decomposed in the multiplets that are natural to the KK black hole.}
	\label{KKres}
\end{table}

\medskip
\begin{table}[h]
\centering
\begin{tabular}{|c||c|c|c|c|c|}
\hline
Multiplet / Theory & ${\cal N}=8$ & ${\cal N}=6$  & ${\cal N}=4$ & ${\cal N}=2$ & ${\cal N}=0$ \cr
\hline
\hline
{\rm KK block} & $1$ &  $1$ & $1$ & $1$ & $1$\cr
\hline
{\rm Gravitino block} & $4$ & $3$ & $2$ & $1$ & $0$ \cr
\hline
{\rm Vector block}  & $27$ & $15$ & $n+5$ & $n_V$ & $0$ \cr
\hline
{\rm Gaugino block} &  $24$ & $10$  & $2n$ & $n_V+n_H-1$ & $0$ \cr
\hline
{\rm Scalar block} & $42$  & $14$  & $5n-4$ & $n_V+4n_H-1$ & $0$ \cr
\hline
$a$ & $5\/2$ &$3\/2$&${1\/32}(22+3n)$&$ {1\/192}(65+17 n_V + n_H)$&$31\/72$\cr
\hline
$c$ & $0$ &$0$&${3\/32}(2+n)$&$ {3\/64} (17+n_V+n_H )$&$37\/24$\cr
\hline
\end{tabular}
\caption{The degeneracy of multiplets in the spectrum of quadratic fluctuations around the KK black hole embedded in to various theories, and their respective values of the $c$ and $a$ coefficients defined in \eqref{eq:a4min}.  For ${\cal N}=4$, the integer $n$
	is the number of ${\cal N} =  4$ matter multiplets. For ${\cal N}=2$, the recorded values of $c$ and $a$ for the gravitino and the gaugino blocks were
	only established for $ST(n_V-1)$ models. }
\label{tab:degeneracy1}
\end{table}

It is worth making a few remarks.
\begin{enumerate}
\item The value of $c-a$ in each case is independent of the couplings of the theory. In other words, $c-a$ can be reproduced by an equal number of minimally
coupled fields on the same black hole background. This property is due to the fact that none of the non-minimal couplings appearing in our blocks involve
 the Riemann tensor $R_{\mn\rs}$. Therefore, the coefficient of $R_{\mn\rs}R^{\mn\rs}$ is insensitive to the non-trivial couplings.

\item The values of $c$ for blocks recorded in Table \ref{KKres} do not have any obvious regularity, they are not suggestive of any cancellations.
The vanishing of the $c$-anomaly for the $\cN=6$ and $\cN=8$ theories, exhibited in Table  \ref{tab:degeneracy1}, seems therefore rather miraculous.
Somehow these embeddings with large supersymmetry have special properties that are not shared by those with lower supersymmetry.
\end{enumerate}

\subsection{Quantum Corrections to Black Hole Entropy}\label{sec:bh}

The logarithmic terms in the one-loop effective action of the massless modes correct the entropy of the black hole as
\be\label{eq:sbhlog}
\delta S_{\rm BH} = {1\over 2}(C_{\rm local} +C_{\rm zm}) \log {A_H\over G}~.
\ee
In this subsection we gather our results and evaluate the quantum contribution for the Kaluza-Klein black hole.

The local contribution is given by the integrated form of the Seeley-DeWitt coefficient $a_4(x)$:
\be\label{eq:clocalbh}
C_{\text{local}}= {c\/16\pi^2} \int \sqrt{g} \,d^4x \, W_{\mn\rs}W^{\mn\rs}- {a \/16\pi^2}  \int \sqrt{g} \,d^4x \, E_4~.
\ee
The second term is essentially the Euler characteristic
\be
\chi = {1\/32 \pi^2}\int d^4x  \sqrt{g} \,E_4 =2~,
\ee
for any non-extremal black hole. It is a topological invariant so it does not depend on black hole parameters. In contrast, the first integral in (\ref{eq:clocalbh})
depends sensitively on the details of the black hole background. Using the KK black hole presented in section \ref{sec:KK theory} with $J=0$ and $P=Q$ we
find
\be\label{eq:intweylKK}
{1\/16 \pi^2}\int d^4x  \sqrt{g} \, W_{\mn\rs}W^{\mn\rs} = 4 + {8\/ 5\, \xi(1+\xi)}~,
\ee
where $\xi\geq 0$ is a dimensionless parameter related to the black hole parameters as
\be
{Q\over G M} ={P\over G M} = { \sqrt{2(1+\xi)} \/ 2+\xi} ~.
\label{eqn:QMmu}
\ee
In this parametrization the extremal (zero temperature) limit corresponds to $\xi\to 0$ and the Schwarzschild (no charge) limit
corresponds to $\xi\to\infty$.

We also need to review the computation of $C_{\text{zm}}$, the integer that captures corrections to the effective action due to zero modes.
In our schematic notation zero modes $\lambda_i=0$ are included in the heat kernel (\ref{eq:dsl}) and therefore contribute to the local term $C_{\text{local}}$.
However, the zero mode contribution to the effective action is not computed correctly by the Gaussian path integral implied in (\ref{eqn:effaction}) and
should instead be replaced by an overall volume of the symmetry group responsible for the zero mode. It is the combination of removing the zero-mode from
the heat kernel and adding it back in again as a volume factor that gives the correction $C_{\text{zm}}$.

Additionally, the effective action defined by the Euclidean path integral with thermal boundary conditions is identified with the free energy in the canonical ensemble whereas the entropy is computed in the microcanonical ensemble where mass and charges are fixed. The Legendre transform relating these ensembles gives
a logarithmic contribution to the entropy that we have absorbed into $C_{\text{zm}}$, for brevity. 

The various contributions to $C_{\text{zm}}$ are not new, they were analyzed in \cite{Sen:2012dw}. The result can be consolidated
in the formula \cite{Charles:2015eha}
\be \label{czm}
C_{\text{zm}} = -(3+K)+2N_{\SUSY}+3\,\delta_{\text{non-ext}}~.
\ee
Here $K$ is the number of rotational isometries of the black hole, $N_{\SUSY}$ is the number of preserved real supercharges. $\delta_{\text{non-ext}}$ is $0$ if
the black hole is extremal and $1$ otherwise. The non-extremal KK black hole with $J=0$ is spherically symmetric and
has $K=3$, $N_{\text{SUSY}} = 0$ and $\d_{\text{non-ext}} = 1$. Therefore, $C_{\text{zm}} = -3$ for all the non-extremal black holes we consider in this paper but
$C_{\text{zm}} = -6$ in the extreme limit.

Combining all contributions, our final result for the coefficient of the logarithmic correction to the non-extreme black hole entropy is
\be\label{eq:finalfinal}
 {1\/2} (C_{\text{local}} + C_{\text{zm}}) =  2 (c-a) -{3\/2} +  {4\/5 \,\xi(1+\xi)}\,c~,
\ee
where the values of $c$ and $a$ for the theories discussed in this paper are given in Table \ref{tab:degeneracy1}.
The expression manifestly shows that when $c\neq 0$, which is the case for ${\cal N}=0,2,4$, the quantum correction to the entropy depends on
black hole parameters through $\xi$ or, by the relation (\ref{eqn:QMmu}), through the physical ratio ${Q/G M}$.
The cases with very high supersymmetry are special since $c=0$ when ${\cal N}\geq 6$ and then the coefficient of the logarithm is purely numerical.
For example, we find the quantum corrections
\be
\delta S^{({\cal N}=6)}_{\textrm{non-ext}} = - {9\over 2}\log {A_H\over G}~, \qquad \delta S^{({\cal N}=8)}_{\textrm{non-ext}} = - {13\over 2}\log {A_H\over G}~,
\ee
to the non-extremal black holes on the non-BPS branch.

As we have stressed, the KK black hole on the non-BPS branch is not intrinsically exceptional. In the non-rotating case with $P=Q$ that is our primary focus,
the geometry is the standard Reissner-Nordstr\"{o}m black hole. However, Kaluza-Klein theory includes a scalar field, the dilaton, and this dilaton couples
non-minimally to gravity and to the gauge field. According to Table \ref{tab:degeneracy1} we find $c = {37\over 24}$ for the KK black hole that is, after all,
motivated by a higher dimensional origin.

An appropriate benchmark for this result is the minimally coupled Einstein-Maxwell theory, which has Reissner-Nordstr\"{o}m as a solution, with an additional minimally coupled scalar field. The KK theory and the minimal theory both have $c-a={10\over 9}$, because these theories have the same field content,
and the zero-mode content of the black holes in the two theories is also identical, because the geometries are the same. However,
$c={55\over 24}$ for the minimally coupled black hole, a departure from the KK black holes. Thus, as one would expect, the quantum corrections to the black
hole entropy depend not only on the field content but also on the couplings to low energy matter.

Although the focus in this paper has been on the non-extreme case, and specifically whether the logarithmic corrections to the black hole entropy
depend on the departure from extremality, it is worth highlighting the extremal limit since in this special case a detailed microscopic model is the
most realistic. In the extremal case we find the quantum correction on the non-BPS branch
\be
\delta S_{\textrm{ext}} = - {\cal N} \log {A_H\over G}~,
\ee
for ${\cal N}=6,8$. The surprising simplicity of this result is inspiring.

\section{Discussion}\label{sec:discussion}

In summary, we have shown that the spectrum of quadratic fluctuations around static Kaluza-Klein black holes in four dimensional supergravity partially
diagonalizes into blocks of fields. Tables \ref{KKres} and \ref{tab:degeneracy1} give the $c$ and $a$ coefficients
that control the Seeley-DeWitt coefficient $a_4(x)$ for each block and, taking into account appropriate degeneracies, for
each supergravity theory.  These coefficients directly yield the logarithmic correction to the black hole entropy via (\ref{eq:sbhlog}-\ref{eq:clocalbh}).

The detailed computations are quite delicate since any improper sign or normalization can dramatically change our conclusions.
We therefore proceeded with extreme care, devoting several sections to explain the embedding of the Kaluza-Klein black hole into a
range of supergravities and carefully record the action for quadratic fluctuations of the fields around the background. Moreover, we allowed for
considerable redundancy, with indirect symmetry arguments supporting explicit computations and also performing many computations both
analytically and using Mathematica. These steps increase our confidence in the results we report.

The prospect that interesting patterns in these corrections could lead to novel insights into black hole microstates is our main motivation for computing these
quantum corrections in supergravity theories. Our discovery that $c=0$ for ${\cal N}=6,8$ on the non-BPS branch is therefore gratifying.
Recall that when $c$ vanishes, the quantum correction is universal, it depends on the matter content of the theory but not on the parameters of the black hole.
This property therefore holds out promise for a detailed microscopic description of these corrections. Such progress would be welcome since our current understanding of, for example, the $D0-D6$ system leaves much to be
desired \cite{Larsen:1999pu,EmparanHorowitz2006,GimonLarsenSimon2008,GimonLarsenSimon2009} for the non-BPS branch.

Conversely, our analysis shows that on the non-BPS branch $c\neq 0$ for ${\cal N}\leq4$. On the BPS-branch not only has it been found that
$c=0$ for all ${\cal N}\geq2$ but this fact has also been shown to be a consequence of ${\cal N}=2$ supersymmetry \cite{Charles2017}. It would
be interesting to similarly understand why $c=0$ requires ${\cal N}\geq6$ on the non-BPS branch.

To date, there is no known microstate counting formula that, when compared to the black hole entropy, accounts for terms that involve $c\neq 0$. For example, in {\it all} cases considered in \cite{Mandal:2010cj,Sen2014,Belin:2016knb}, the object of interest is an index, or a closely related avatar, and the resulting logarithmic terms
nicely accommodate quantum corrections when $C_{\rm local}$ is controlled by $a$ alone. The challenge of reproducing the logarithmic correction when $c$ is non-vanishing comes from the intricate dependence  on the black hole parameters that the Weyl tensor gives to $C_{\rm local}$.   It would be interesting to
understand which properties a partition function must possess in order that the logarithmic correction to the thermodynamic limit leads to $c\neq 0$.

An interesting concrete generalization of the present work would be to increase the scope of theories considered. In section \ref{sec:N=2sugra} our main
obstacle to covering all ${\cal N}=2$ theories is the complicated structure of fermion couplings for a generic prepotential, and hence we restrict the discussion
in section \ref{N=2:bosons} to the $ST(n)$ models.  Nevertheless, we suspect that for a generic prepotential our conclusions would not be significantly different. In particular, we predict that $c\neq 0$ on the non-BPS branch for any ${\cal N}=2$ supergravity. It would of course be desirable to confirm this explicitly.

A more ambitious generalization would be to consider more general black hole solutions, specifi\-cally those where the dilaton $\Phi^{(\text{KK})}$ is not constant.
Our assumption that $\Phi^{(\text{KK})}=0$ simplified our computations greatly by sorting quadratic fluctuations into blocks that are decoupled from one another.
By addressing the technical complications due to relaxation of this assumption and so computing $a_4(x)$ for black holes with non-trivial dilaton we could, in particular,
access solutions with non-zero angular momentum $J\neq0$. The rotating black holes on the non-BPS branch are novel since they never have constant
dilaton, even in the extremal limit \cite{AstefaneseiGoldsteinJenaEtAl2006}. Therefore, they offer an interesting contrast to the Kerr-Newman black hole,
their counterparts on the BPS branch \cite{Charles:2015eha}. Rotation is quite sensitive to microscopic details so any differences or similarities between the
quantum corrections to rotating black holes on the BPS and non-BPS branches may well provide valuable clues towards a comprehensive
microscopic model. A nonconstant dilaton is also the linchpin to connections with the new developments in AdS$_2$ holography for rotating black holes such as in  \cite{AnninosAnousDAgnolo2017,CLP}.

\section*{Acknowledgements}
We thank Anthony Charles and Garrett Goon for useful discussions. This work was supported in part by the U.S. Department of Energy under grant DE-FG02-95ER40899. AC is supported by Nederlandse Organisatie voor Wetenschappelijk Onderzoek (NWO) via a Vidi grant. The work of AC and VG is part of the Delta ITP consortium, a program of the NWO that is funded by the Dutch Ministry of Education, Culture and Science (OCW).


\newpage
\appendix
\section{Computations of Seeley-DeWitt Coefficients}
\label{app:a4}

In this appendix, we give the details on the computation of the Seeley-DeWitt coefficients for Kaluza Klein black holes and their embeddings in ${\cal N}\geq2$
supergravity. Most of the computations were done using the Mathematica package \ttt{xAct}. We present our results according to the organization of quadratic fluctuations into blocks that was introduced in section \ref{sec:fluctuations}.

\medskip

The basic steps of our implementation are:
\begin{enumerate}
	\item We expand the Lagrangian to second order.\footnote{For fermions we always write the quadratic fluctuations with Majorana spinors, following the conventions of \cite{Charles:2015eha}. }  This was done in sections \ref{sec:fluctuations} and \ref{sec:N=2sugra} for the supergravity theories of interest. The bosonic Lagrangian can also be expanded using \ttt{xPert}.
	\item We gauge-fix and add the corresponding ghosts. The gauge-fixing and the ghosts were detailed for each block in sections \ref{sec:fluctuations} and \ref{sec:N=2sugra}. In this appendix, we highlight and record their contributions to the heat kernel.
	\item We rearrange the fluctuation operator $\L_m^n$ so that it takes the canonical form \eqref{eq:diff}. We then read off the operators $\omega_\mu$ and $P$
	and compute the operators $E$ and $\Om_\mn$. These are the most cumbersome steps so they are executed primarily using Mathematica. Since some expressions are rather lengthy for the matrix operators due to the non-minimal couplings, we mostly present the traces of these operators.	
	\item We compute the Seeley-DeWitt coefficient $a_4(x)$ using formula (\ref{a4}). This also includes the ghosts from the second step.
	\item We simplify $a_4(x)$ using the equations of motion, tensor and gamma matrix identities. This brings $a_4(x)$ to its minimal form \eqref{eq:a4min}, where we can read off the coefficients $c$ and $a$.
\end{enumerate}

\subsection{Preliminaries}\label{app:pre}

We use the following formula to compute the Seeley-DeWitt  coefficient
\be\label{a41}
(4\pi)^2 a_4(x) = \Tr\left[{1\over2} E^2 + {1\over6} R E + {1\over 12} \Om_\mn\Om^\mn + {1\over360} (5 R^2 + 2R_{\mn\rs}R^{\mn\rs} - 2R_\mn R^\mn)\right]~.
\ee
This object further simplifies due to the equations of motion, Bianchi, and Schouten identities.  These simplifications imply that we can cast \eqref{a41} in the form
\be\label{a42}
a_4(x)= {c\over 16\pi^2} W_{\mu\nu\rho\sigma}W^{\mu\nu\rho\sigma}-{a\over 16\pi^2} E_4~,
\ee
where the square of the Weyl tensor is
\be\label{eq:w2}
W_{\mu\nu\rho\sigma}W^{\mu\nu\rho\sigma}= R_{\mu\nu\rho\sigma}R^{\mu\nu\rho\sigma}-2R_{\mu\nu}R^{\mu\nu}+{1\over 3}R^2~,
\ee
and the Euler density is
\be\label{eq:e4}
E_4=R_{\mu\nu\rho\sigma}R^{\mu\nu\rho\sigma}-4R_{\mu\nu}R^{\mu\nu}+R^2~.
\ee

For each block, as summarized in Table \ref{FluctTable}, we will report both \eqref{a41} and \eqref{a42}. The identities used to simplify \eqref{a41} to its minimal form \eqref{a42} are listed below. For fermionic fluctuations, we also use many gamma matrix identities which are well known and not repeated here.\\

\noindent\textbf{On-shell conditions:} The equations of motion background with constant dilaton are
	\begin{align}\label{eq:s1}
	F_{\mu \alpha}F_\nu^{\' \alpha}  =2 R_{\mu \nu}~, \qquad R= 0~,\\\notag
	F_\mn F^\mn = 0~,\qquad
	D_\mu F^{\mu \nu} = 0~.
	\end{align}
\noindent\textbf{Bianchi identities:}  Starting from
\be
	\nabla_\mu \tF^{\mu \nu} = 0~, \qquad R_{\mu [\nu \alpha \beta]}=0~,
\ee
where $ \tF_{\mu \nu} =-{i\over 2}\epsilon_{\mu \nu \alpha\beta}F^{\alpha\beta}$ we find
	\begin{align}\label{eq:s2}
	R_{\mu \nu \alpha \beta}R^{\mu \alpha \nu \beta} &={1\/2}R_{\mu \nu \alpha \beta}R^{\mu \nu \alpha \beta}~, \\\notag
	(D_\alpha F_{\mu \nu})(D^\nu F^{\mu \alpha}) &={1\/2}(D_\alpha F_{\mu \nu})(D^\alpha F^{\mu \nu})~, \\\notag
	F^{\alpha\nu}(D_\alpha F_{\mu \nu})  &= {1\/2} F^{\nu \alpha}(D_\mu F_{\nu \alpha})~, \\\notag
	R_{\mu \alpha \nu \beta} F^{\mu \nu} F^{\alpha \beta} &= {1\/2} R_{\mu \nu \alpha \beta} F^{\mu \nu} F^{\alpha \beta}~, \\\notag
	\epsilon_{\mu \nu \alpha \beta} D^\alpha F^{\rho \beta} &= {1\/2} \epsilon_{\mu \nu \alpha \beta} D^\rho F^{\alpha \beta}~.
	\end{align}

\noindent\textbf{Schouten identities:} The Schouten identity is $g^{\mu[\nu} \epsilon^{\rs\tau\la]} = 0$. From this, we can derive
	\begin{align}\label{eq:s3}
	\tF_{\mu \alpha}F_\nu^{\'\alpha} &= {1\/4} g_{\mu \nu} \tF_{\alpha \beta}F^{\alpha \beta}
	\end{align}
\noindent\textbf{Derivative relations:} The following identity is also useful
	\begin{align}
(D_\alpha F_{\mu \nu} )(D^\alpha F^{\mu \nu} )&=- 2 R_{\mu \nu} F^{\mu \alpha} F^\nu_{\'\alpha} + R_{\mu \nu \alpha \beta} F^{\mu \nu} F^{\alpha \beta}
	\end{align}
and holds up to a total derivative.


\subsection{KK Block}

The quadratic Lagrangian is given in (\ref{N=8:CMb}). To evaluate the Seeley-DeWitt coefficient, the kinetic term of $h_{\mu\nu}$ is analytically continued to
\be
h^{\text{new}}_{\mu\nu} = -{i\over 2}h_{\mu\nu}~,
\ee
for the kinetic term to have the right sign. In addition, in order to project onto the traceless part of a symmetric tensor, we define
 \be
G_\rs^{\mn} ={1\/2} \left(\d^\mu_{\'\rho} \d^\nu_{\'\s} + \d^\mu_{\'\s}\d^\nu_{\'\rho} - {1\/2}g^\mn g_\rs \right)~.
\ee
Traces of operators must be taken after contraction with this tensor. For example, for a four index operator $O$ we use
\be
\Tr \,O= G_\rs^{\mn} O^\rs_{\mn}~.
\ee

The relevant  traces that appear in \eqref{a41} for the KK block are
\begin{eqnarray}\label{eq:gravitontr}
\Tr\,E &=&\; 3 F_\mn F^\mn -  7 R ~,\\\notag
\Tr\, E^2 &=& \;{33\/16} F^\mu_{\'\rho} F^{\nu\rho} F_{\mu\s} F_\nu^{\'\s} + {21\/16} F_\mn F^\mn F_\rs F^\rs - 5 R_\mn R^\mn - {5\/2} R_\mn F^{\mu}_{\'\rho}F^{\nu\rho}  -{1\/2} R F_\mn F^\mn \\\notag
&& + 5 R^2 + 2R_{\mn\rs}R^{\mn\rs} + 2 R_{\mu \rho\nu\s} R^{\mn\rs} - 2 F^{\mn}_{\'\'\; ;\mu} F_{\nu\';\rho}^{\'\rho} + {1\/2}F_{\mu\rho;\nu}F^{\mn;\rho}+{1\/2} F_{\mn;\rho}F^{\mn;\rho}~,\\\notag
\Tr\,\Om_\mn\Om^\mn &=& -{7\/8}F^\mu_{\'\rho} F^{\nu\rho} F_{\mu\s} F_\nu^{\'\s} -{23\/8}F_\mn F^\mn F_\rs F^\rs + 2 R_\mn F^{\mu}_{\'\rho}F^{\nu\rho} +R F_\mn F^\mn  \\\notag
&& + 3 R_{\mu\rho\nu\s}F^\mn F^\rs - 7 R_{\mn\rs}R^{\mn\rs} - F^\mn_{\'\'\; ;\mu} F_{\nu\';\rho}^{\'\rho} + 4 F_{\mu\rho;\nu}F^{\mn;\rho}- 8 F_{\mn;\rho}F^{\mn;\rho}~.
\end{eqnarray}

The gauge-fixing also introduces ghosts with the Lagrangian
\be
e^{-1}\cL_{\text{ghosts}} = 2 b_\mu \left(\Box g^\mn+R^\mn \right)c_\nu + 2 b\Box c - 4 b F^\mn D_\mu c_\nu~,
\ee
where $b_\mu,c_\mu$ are vector ghosts associated to the graviton and $b,c$ are scalar ghosts associated to the graviphoton.
The contribution of the ghosts are
\begin{eqnarray}\label{eq:ghosttr}
\Tr\,E &=& 2 R ~, \\\notag
\Tr\,E^2 &=& 2 R_\mn R^\mn~,\\\notag
\Tr\, \Om_\mn\Om ^\mn &=& -2 R_{\mn\rs}R^{\mn\rs}~.
\end{eqnarray}
The total ghost contribution is
\be\label{eq:ghosttotal}
(4\pi)^2 a^{\rm ghost}_4(x)= {1\over 9} R_{\mn\rs}R^{\mn\rs} - {17\over 18} R_\mn R^\mn - {17\over 36} R^2~.
\ee

Combining the contributions \eqref{eq:gravitontr} and \eqref{eq:ghosttotal} gives
\begin{eqnarray}
(4\pi)^2 a_4(x) &=& \; {23\/24} F^\mu_{\'\rho} F^{\nu\rho} F_{\mu\s} F_\nu^{\'\s} + {5\/12} F_\mn F^\mn F_\rs F^\rs - {127\/36} R_\mn R^\mn - {13\/12} R_\mn F^{\mu}_{\'\rho}F^{\nu\rho}  \\\notag
&& +{1\/3} R F_\mn F^\mn  + {77\/72} R^2 + {1\/4}R_{\mu\rho\nu \s}F^\mn F^\rs + {11\/18}R_{\mn\rs}R^{\mn\rs} + R_{\mu \rho\nu\s} R^{\mn\rs} \\\notag
&& - {13\/12} F^{\mn}_{\'\'\; ;\mu} F_{\nu\';\rho}^{\'\rho} + {7\/12}F_{\mu\rho;\nu}F^{\mn;\rho}-{5\/12} F_{\mn;\rho}F^{\mn;\rho}~.
\end{eqnarray}
We use the identities listed in (\ref{eq:s1}-\ref{eq:s3}) to obtain
\be
(4\pi)^2 a_4(x) = {10\/9} R_{\mn\rs}R^{\mn\rs}  - {49\/36} R_\mn R^\mn ~,
\ee
and from here we find
\be
a_{\rm KK} = {31\/72}~,\qquad c_{\rm KK}={37\/24}~.
\ee

\subsection{Vector Block}

The vector block in its minimal form is described by the quadratic Lagrangian \eqref{N=2:vector} and for the matter content of ${\cal N}=8$ by (\ref{N=8:vector}).
The matrices that appear in the quadratic fluctuation operator are
\be\nonumber
E  =\le( \begin{matrix} {1\/4} F_\mu^{\'\rho}F_{\nu\rho}  - R_\mn &  {1\/2} F_{\nu\';\rho}^{\'\rho}\\
{1\/2} F_{\mu\';\rho}^{\'\rho}& -{1\/4} F_\rs F^\rs \end{matrix}\ri)~, \quad
\Om_\rs =\le( \begin{matrix}  R_{\mn\rs}  + {1\/4} F_{\mu\s} F_{\nu\rho}-{1\/4} F_{\mu\rho} F_{\nu\s} & \quad {1\/2}F_{\mu\s;\rho}-{1\/2} F_{\mu\rho;\s} \\
-{1\/2}F_{\nu\s;\rho}+{1\/2} F_{\nu\rho;\s}  & 0 \end{matrix}\ri)~,
\ee
where the first row/column corresponds to the vector field and the second row/column to the scalar field. The relevant traces are
\begin{eqnarray}
\Tr\, E &=& -R~, \\
\Tr\, E^2 &=& {1\/16}F^\mu_{\'\rho} F^{\nu\rho} F_{\mu\s} F_\nu^{\'\s} +{1\/16} F_\mn F^\mn F_\rs F^\rs + R_\mn R^\mn \\\notag
&&- {1\/2}R_\mn F^{\mu}_{\'\rho}F^{\nu\rho}  - {1\/2}F^{\mn}_{\'\'\;;\mu} F_{\nu\'\,;\rho}^{\'\rho} ~,\\
\Tr\, \Om_\mn \Om^\mn &=& {1\/8}F^\mu_{\'\rho} F^{\nu\rho} F_{\mu\s} F_\nu^{\'\s} -{1\/8} F_\mn F^\mn F_\rs F^\rs +  R_{\mu \rho\nu\s}F^\mn F^\rs\\\notag
&& -  R_{\mn\rs}R^{\mn\rs} +  F_{\mu\rho;\nu}F^{\mn;\rho}-  F_{\mn;\rho}F^{\mn;\rho}~.
\end{eqnarray}

The ghosts for the vector block are two minimally coupled scalars with fermionic statistics. Their contribution to the Seeley-DeWitt coefficient is
\be
(4\pi)^2 a^{\rm ghost}_4(x)= -{1\over 180}(2R_{\mn\rs}R^{\mn\rs}-2R_{\mn}R^{\mn} + 5R^2)~.
\ee

We combine the contributions of the vector block and its associated ghosts and get
\begin{eqnarray}
(4\pi)^2 a_4(x) &=& {1\/24}F^\mu_{\'\rho} F^{\nu\rho} F_{\mu\s} F_\nu^{\'\s} +{1\/48} F_\mn F^\mn F_\rs F^\rs +{29\/60} R_\mn R^\mn \\\notag
&&- {1\/4}R_\mn F^{\mu}_{\'\rho}F^{\nu\rho}  - {1\/8}R^2+ {1\/12} R_{\mu \rho\nu\s}F^\mn F^\rs -{1\/15} R_{\mn\rs}R^{\mn\rs} \\\notag
&& -{1\/4}F^{\mn}_{\'\'\;;\mu} F_{\nu\'\,;\rho}^{\'\rho} + {1\/12}F_{\mu\rho;\nu}F^{\mn;\rho}- {1\/12} F_{\mn;\rho}F^{\mn;\rho}
\end{eqnarray}
After using the identities (\ref{eq:s1}-\ref{eq:s3}), we obtain
\be
(4\pi)^2 a_4(x) = -{1\/15} R_{\mn\rs}R^{\mn\rs}  +{19\/60} R_\mn R^\mn~.
\ee
This leads to
\be\label{eq:avector}
a_{\rm vector} = {11\/120}~, \qquad c_{\rm vector}= {1\/40}~.
\ee
When the vector block contains a pseudoscalar instead of a scalar, such as in \eqref{N=8:Pvector}, the result remains the same because of simplifications due to our background.

\subsection{Gravitino Block}

The gravitino block is characterized by the quadratic Lagrangian (\ref{N=8:gravitino}).  After using gamma matrix identities, the relevant traces are
\begin{eqnarray}\label{eq:gravitinoblock}
\Tr\, E &=& {1\/2} F_\mn F^\mn + {1\/2}\tF_\mn\tF^\mn-10 R ~,\\\notag \\
\Tr\,E^2 &=& -{105\/128}F^\mu_{\'\rho} F^{\nu\rho} F_{\mu\s} F_\nu^{\'\s} +{81\/128} F_\mn F^\mn F_\rs F^\rs + {43\/64} F^\mn F^\rs \tF_{\mu\rho}\tF_{\nu\s}  \\\notag
&& - {13\/32} F^{\mu}_{\'\rho} F^{\nu\rho} \tF_{\mu}^{\'\s} \tF_{\nu\s} + {7\/128} \tF^{\mu}_{\'\rho} \tF^{\nu\rho} \tF_{\mu}^{\'\s} \tF_{\nu\s}  - {21\/64 }F_\mn F^\mn \tF_{\rs}\tF^\rs  \\\notag
&&+ {9\/128}\tF_\mn \tF^\mn \tF_{\rs}\tF^\rs  -{1\/4} R F_\mn F^\mn -{1\/4} R \tF_\mn \tF^\mn + {5\/2} R^2 \\\notag
&& -{3\/2} R_{\mu\rho\nu\s}F^\mn F^\rs+{3\/2}R_{\mu\rho\nu\s}\tF^\mn \tF^\rs + 4R_{\mn\rs}R^{\mn\rs} \\\notag
&& -{7\/2}F_{\mu\rho;\nu}F^{\mn;\rho}+3 F_{\mn;\rho}F^{\mn;\rho}  +{3\/2}\tF_{\mu\rho;\nu}\tF^{\mn;\rho}-2 \tF_{\mn;\rho}\tF^{\mn;\rho} ~,\\ \notag \\
\Tr\, \Om_\mn \Om^\mn &=&{185\/64}F^\mu_{\'\rho} F^{\nu\rho} F_{\mu\s} F_\nu^{\'\s} -{185\/64} F_\mn F^\mn F_\rs F^\rs - {27\/32} F^\mn F^\rs \tF_{\mu\rho}\tF_{\nu\s}   \\\notag
&&  - {3\/16} F^{\mu}_{\'\rho} F^{\nu\rho} \tF_{\mu}^{\'\s} \tF_{\nu\s} + {9\/64} \tF^{\mu}_{\'\rho} \tF^{\nu\rho} \tF_{\mu}^{\'\s} \tF_{\nu\s}  +{33\/32 }F_\mn F^\mn \tF_{\rs}\tF^\rs  \\\notag
&&-{9\/64}\tF_\mn \tF^\mn \tF_{\rs}\tF^\rs  +7 R_{\mu\rho\nu\s}F^\mn F^\rs-{3}R_{\mu\rho\nu\s}\tF^\mn \tF^\rs -13 R_{\mn\rs}R^{\mn\rs} \\\notag
&&+ 7 F_{\mu\rho;\nu}F^{\mn;\rho}- 7 F_{\mn;\rho}F^{\mn;\rho}-3\tF_{\mu\rho;\nu}\tF^{\mn;\rho}+3 \tF_{\mn;\rho}\tF^{\mn;\rho}  ~.
\end{eqnarray}
The gauge-fixing produces fermionic ghosts $b_A,c_A,e_A$ with Lagrangian
\be
e^{-1}\cL_{\text{ghost}} = \bar{b}_A \g^\mu D_\mu c_A + \bar{e}_A \g^\mu D_\mu e_A~,
\ee
where $A=1,2$ is the flavor index. This simply corresponds to six minimally coupled Majorana fermions which contribute with an opposite sign. Their Seeley-DeWitt contribution is
\be\label{eq:ghostgravitino}
(4\pi)^2 a^{\rm ghost}_4(x) =-{1\over 120}\left(7R_{\mn\rs}R^{\mn\rs}  +8 R_\mn R^\mn - 5R^2\right)~.
\ee

Combining \eqref{eq:gravitinoblock} and \eqref{eq:ghostgravitino} gives
\begin{eqnarray}
(4\pi)^2 a_4(x) &=&{65\/768}F^\mu_{\'\rho} F^{\nu\rho} F_{\mu\s} F_\nu^{\'\s} -{29\/768} F_\mn F^\mn F_\rs F^\rs-{17\/128}F^\mn F^\rs \tF_{\mu\rho}\tF_{\nu\s}   \\\notag
&&  + {7\/64} F^{\mu}_{\'\rho} F^{\nu\rho} \tF_{\mu}^{\'\s} \tF_{\nu\s} - {5\/256} \tF^{\mu}_{\'\rho} \tF^{\nu\rho} \tF_{\mu}^{\'\s} \tF_{\nu\s}  +{5\/128 }F_\mn F^\mn \tF_{\rs}\tF^\rs \\\notag
&& -{3\/256}\tF_\mn \tF^\mn \tF_{\rs}\tF^\rs  +{2\/45}R_\mn R^\mn + {1\/48} R F_\mn F^\mn + {1\/48} R \tF_\mn \tF^\mn  \\\notag
&& - {1\/36} R^2   +{1\/12} R_{\mu\rho\nu\s}F^\mn F^\rs - {1\/4}R_{\mu\rho\nu\s}\tF^\mn \tF^\rs -{113\/180} R_{\mn\rs}R^{\mn\rs} \\\notag
&& + {7\/12} F_{\mu\rho;\nu}F^{\mn;\rho}- {11\/24} F_{\mn;\rho}F^{\mn;\rho} -{1\over 4}\tF_{\mu\rho;\nu}\tF^{\mn;\rho}+{3\over 8} \tF_{\mn;\rho}\tF^{\mn;\rho}  ~.
\end{eqnarray}
Using the identities (\ref{eq:s1}-\ref{eq:s3}) gives
\be
(4\pi)^2 a_4(x)  = -{113\/180} R_{\mn\rs}R^{\mn\rs}  + {767\/720 } R_\mn R^\mn~,
\ee
and this leads to
\be
a_{\rm gravitino}=  -{137\/1440},\qquad c_{\rm gravitino} = -{347\/480}~.
\ee

\subsection{Gaugino Block}

The gaugino block is given by  the Lagrangian (\ref{N=8:gaugino}). In this case, the relevant traces are
\begin{eqnarray}
\Tr\, E &=& {1\/4} F_\mn F^\mn - 2 R~,\\
\Tr\, E^2 &=& -{1\/32}F^\mu_{\'\rho} F^{\nu\rho} F_{\mu\s} F_\nu^{\'\s} +{3\/128} F_\mn F^\mn F_\rs F^\rs - {1\/8} R F^\mn F_\mn + {1\/2}R^2 \\\notag
&& - {1\/2}F_{\mu\rho;\nu}F^{\mn;\rho}+{1\/4} F_{\mn;\rho}F^{\mn;\rho}~,\\
\Tr\, \Om_\mn\Om^\mn &=& {1\/8}F^\mu_{\'\rho} F^{\nu\rho} F_{\mu\s} F_\nu^{\'\s} -{1\/8} F_\mn F^\mn F_\rs F^\rs  + R_{\mu\rho\nu\s} F^\mn F^\rs- R_{\mn\rs}R^{\mn\rs}  \\\notag
&&+ F_{\mu\rho;\nu}F^{\mn;\rho}-  F_{\mn;\rho}F^{\mn;\rho}~.
\end{eqnarray}
The Seeley-DeWitt coefficient is
\begin{eqnarray}
(4\pi)^2 a_4(x) &=& {1\/384}F^\mu_{\'\rho} F^{\nu\rho} F_{\mu\s} F_\nu^{\'\s} -{1\/1536} F_\mn F^\mn F_\rs F^\rs +{1\/45} R_\mn R^\mn + {1\/96} R F^\mn F_\mn \\
&& -{1\/72}R^2 - {1\/24} R_{\mu\rho\nu\s}F^\mn F^\rs + {7\/360} R_{\mn\rs}R^{\mn\rs}+ {1\/12}F_{\mu\rho;\nu}F^{\mn;\rho}-{1\/48} F_{\mn;\rho}F^{\mn;\rho}~.\nonumber
\end{eqnarray}
and gives after simplification
\bea
(4\pi)^2 a_4(x) = {7\/360}R_{\mn\rs}R^{\mn\rs} -{ 73\/1440} R_\mn R^\mn~,
\eea
which leads to
\be
a_{\rm gaugino}=-{17\/2880}~, \qquad c_{\rm gaugino}={13\/960}~.
\ee

\bibliographystyle{JHEP-2}
\providecommand{\href}[2]{#2}\begingroup\raggedright\endgroup

\end{document}